\documentclass[12pt]{article}
\usepackage{amsmath}
\usepackage{graphicx,psfrag,epsf}
\usepackage{enumerate}
\usepackage[round]{natbib}
\usepackage{url} 

\usepackage{amsmath,amssymb,amsfonts,amsthm}
\usepackage{float,graphicx, multirow, setspace}
\usepackage{subfiles}
\usepackage{placeins}
\usepackage{url}
\usepackage{hyperref}
\usepackage{xr, xr-hyper}
\usepackage{algpseudocode}
\usepackage{subfig}
\usepackage{algorithm}
\usepackage{makecell}
\usepackage[scr=boondox]{mathalfa}

 \newcommand{\fg}{{\mathsf{FG} }}
 \newcommand{\punt}{{\mathsf{Punt} }}
 \newcommand{\go}{{\mathsf{Go} }}
 
 \newcommand{\bE}{{\mathbb E}}
 \renewcommand{\P}{{\mathbb P}}
 \newcommand{\ind}[1]{\mathbb{I}\left(#1\right)}

 \newcommand{\bx}{{ \mathbf{x}}}

 \newcommand{\ESS}{{ \mathrm{ESS} }}

 \newcommand{\logloss}{{ \mathsf{logloss}}}

\newcommand{\gam}{{ \mathsf{GAM}}}
\newcommand{\xgb}{{ \mathsf{XGBoost}}}

\newcommand{\rf}{{ \mathsf{RF}}}

 \newcommand{\ep}{{ \mathsf{EP}}}

  \newcommand{\tq}{{\mathsf {TQ}}}
 \newcommand{\ps}{{ \mathsf{PS}}}
 \newcommand{\tp}{{ \mathsf{TP}}}
 \renewcommand{\wp}{{ \mathsf{WP}}}
 
 \newcommand{\wpone}{ \mathsf{WP_1} }
 
 \newcommand{\bootp}{{ \mathsf{boot\%} }}
 
 \newcommand{\fgpzero}{{ P_{\mathsf{FG}}^{(0)} }}
 \newcommand{\fgpa}{{ \mathsf{FGPA} }}
 \newcommand{\enyzero}{{  \bE_{\punt}^{(0)}  }}
 \newcommand{\pyoe}{{ \mathsf{PYOE} }}
 \newcommand{\R}{{\mathsf R}}
 \newcommand{\nflfastr}{{\textsf{nflFastR}}}
 \newcommand{\oqot}{{\mathsf {oqot}}}
 \newcommand{\oqdt}{{\mathsf {oqdt}}}
 \newcommand{\dqdt}{{\mathsf {dqdt}}}
 \newcommand{\dqot}{{\mathsf {dqot}}}

 \newcommand{\kq}{{\mathsf {kq}}}
 \newcommand{\pq}{{\mathsf {pq}}}
 \newcommand{\mtry}{{ \mathsf{mtry}}}
 \newcommand{\nodesize}{{ \mathsf{nodesize}}}

\usepackage[dvipsnames]{xcolor}

\newcommand{\blind}{0}

\begin{document}

\bibliographystyle{apalike}

\def\spacingset#1{\renewcommand{\baselinestretch}%
{#1}\small\normalsize} \spacingset{1}


\if0\blind
{
  \title{\bf Analytics, have some humility: a statistical view of fourth-down decision making}
  \author{Ryan S. Brill\thanks{Graduate Group in Applied Mathematics and Computational Science, University of Pennsylvania. Correspondence to: ryguy123@sas.upenn.edu}, \ Ronald Yurko\thanks{Dept.~of Statistics and Data Science, Carnegie Mellon University}, \ and Abraham J. Wyner\thanks{Dept.~of Statistics and Data Science, The Wharton School, University of Pennsylvania}}
  \maketitle
} \fi

\if1\blind
{
  \bigskip
  \bigskip
  \bigskip
  \begin{center}
    {\LARGE\bf Analytics, have some humility: a statistical view of fourth-down decision making}
  \end{center}
  \medskip
} \fi

\bigskip
\begin{abstract}
The standard mathematical approach to fourth-down decision making in American football is to make the decision that maximizes estimated win probability. Win probability estimates arise from machine learning models fit from historical data. These models attempt to capture a nuanced relationship between a noisy binary outcome variable and game-state variables replete with interactions and non-linearities from a finite dataset of just a few thousand games. Thus, it is imperative to knit uncertainty quantification into the fourth-down decision procedure; we do so using bootstrapping. We find that uncertainty in the estimated optimal fourth-down decision is far greater than that currently expressed by sports analysts in popular sports media. 

\end{abstract}


\noindent%
\vfill

\newpage
\spacingset{1.45} 



\section{Introduction}\label{sec:intro}

In-game strategic decision making is one of the fundamental objectives of sports analytics.
To mathematically compare strategies, analysts need a value function that measures the value of each game-state.
The optimal decision maximizes the value of the next game-state.
Across sports, however, value functions are not observable quantities; they are defined by models. 
It is the sports analysts' task to infer the value of each game-state from the massive dataset of all plays in the recent history of a given sport. 

The two most widely used value functions by analysts of American football are win probability and expected points \citep{nflWar}.
\textit{Win probability} ($\wp$) measures the probability that the team with possession at the current game-state wins the game.
\textit{Expected points} ($\ep$) measures the expected value of the net number of points of the next score in the game, relative to the team with possession, given the current game-state.\
The most prominent example of analysts using these value functions to dictate in-game strategy is fourth-down decision making.
On fourth down, a football coach has three choices: go for it ($\go$), attempt a field goal ($\fg$), or punt the ball ($\punt$).
Previous work by \citet{Romer06dofirms} and \citet{Burke4thDownPt3} suggest making the decision that maximizes estimated $\ep$.
Yet a team's goal is to win the game, not score more points on average, so $\ep$ is the wrong objective function. 
Hence, modern approaches by \citet{BaldwinWP} and Burke\footnote{
    Burke releases fourth-down decision recommendations with the ESPN analytics team, and their recommendation algorithm is proprietary.
} suggest making the decision that maximizes estimated $\wp$.
Each of these analyses found that National Football League (NFL) coaches are too conservative on fourth down; they often settle for kicks even when they should go for it. 


The win probability estimates used for fourth-down decision making typically arise from statistical models fit from historical data. 
Given the play-by-play results of the entire recent history of football, these models fit the relationship between a binary win/loss outcome variable and certain game-state variables using data-driven regression or machine learning approaches.
Analysts then recommend the optimal decision according to the model.
This approach overlooks the uncertainty inherent in estimating win probability with limited data.
The win/loss outcome variable is noisy and there are only a few thousand games in the dataset that produce an outcome.
The outcome variable, moreover, is not independent across plays because every non-tied game has only one winner, which reduces the effective sample size of the dataset.
Therefore, it is critical to quantify uncertainty in win probability estimates.
This uncertainty should percolate into the fourth-down decision procedure.


Our focus is not to ``fix'' win probability models by adjusting for additional covariates or reducing model bias.
Rather, we shed light on the high variance nature of estimating win probability and show that such estimates are subject to nontrivial uncertainty.
We use bootstrapping to quantify uncertainty in fourth-down recommendations and recommend a decision when we are confident it has higher $\wp$ than all other decisions.
We find that far fewer fourth-down decisions are as obvious as previously thought.
Thus, we ask football analysts to have some humility: for many game-states, there is simply not enough data to trust $\wp$ point estimates.


The remainder of this paper is organized as follows.
In Section~\ref{sec:wp} we estimate win probability and detail the traditional fourth-down decision procedure. 
In Section~\ref{sec:uncertainty_sec} we knit uncertainty quantification into the decision procedure.
In Section~\ref{sec:fourth_down_dec} we present our main findings and we conclude in Section~\ref{sec:discussion}.

\section{The traditional fourth-down decision procedure}\label{sec:wp}

The standard fourth-down decision procedure involves making the decision that maximizes estimated win probability. 
In this section we overview this procedure and detail how analysts commonly estimate win probability today.
We begin with a brief overview of our historical dataset of football plays in Section~\ref{sec:data}.
Then in Section~\ref{sec:trad_examples} we illustrate the traditional decision process through example plays.
In Section~\ref{sec:fourthd_wp} we estimate the win probability of a fourth-down decision.
These estimates are functions of first-down win probability and decision transition probabilities, which we estimate in Sections~\ref{sec:existing_first_down_wp_models} and \ref{sec:dec_transition_probs}, respectively.

\subsection{Data}\label{sec:data}

We access every NFL play from 1999 to 2022 using the $\R$ package $\nflfastr$ \citep{nflFastR}.
Each play includes variables that describe the context of the play, which are relevant to estimating win probability, such as the score differential, time remaining, yards to opponent endzone (i.e., yardline), down, yards to go, etc. (see Table~\ref{table:wp_variables} in Appendix~\ref{app:data_details} for descriptions of relevant variables).
Note that yards to opponent endzone is an integer in $\{0,1,...,99,100\}$, where $0$ represents a touchdown and $100$ indicates a safety. 
Holding out plays from 1999 to 2005 and from 2022 for various models and validation procedures discussed later in this paper, we are left with a primary dataset of $600,825$ plays, with $229,635$ first-down plays and $4,101$ non-tied games from 2006 to 2021, henceforth referred to as the ``observed play-by-play football dataset.''
We fit our win probability models using this dataset.
The code for this study, which includes code to scrape the dataset used in this study, is publicly available on Github.\footnote{
    \url{https://github.com/snoopryan123/fourth_down}
} 

\subsection{Example plays}\label{sec:trad_examples}

The strength of a traditional decision recommendation is proportional to the effect size, or the estimated gain in win probability by making that decision.
Notably, \citet{BaldwinWP} and Burke employ this decision procedure and post their fourth-down recommendations on X (formerly known as Twitter).\footnote{
    Baldwin's posts his fourth-down recommendations at \texttt{@ben\_bot\_baldwin} and Burke posts his at \texttt{@bburkeESPN}. 
}
Baldwin has posted a recommendation for many fourth-down plays since at least 2021\footnote{
    As of the writing of this paper, \texttt{@ben\_bot\_baldwin} has over 11,000 posts (decision recommendations) on X. 
} and Burke has posted a recommendation for select fourth down plays since at least 2022.
We illustrate this procedure through two example plays.

\begin{figure}[p]
    \centering{}
    {\includegraphics[width=0.575\textwidth]{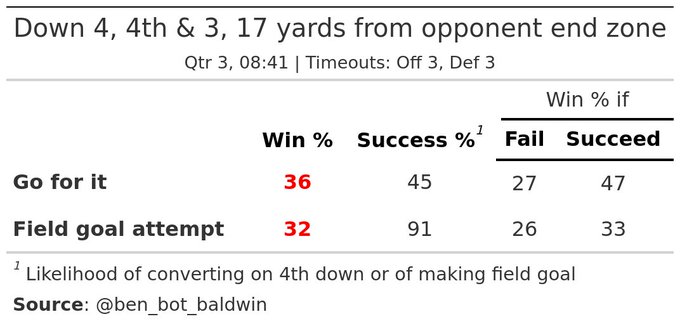} }
    \\
    {\includegraphics[width=0.575\textwidth]{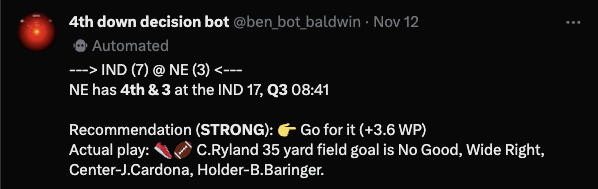} }
    \caption{
        Baldwin's decision charts for example play 1.
        The top image summarizes his models' outputs for each decision, including estimated fourth-down win probability (in red), $\go$/$\fg$ success probability (second column), and win probability given $\go$/$\fg$ failure or success (third and fourth column).
        The bottom image is a screenshot of his post on X, which summarizes the recommended decision and the actual play.
    }
    \label{fig:baldwin_ex}
\end{figure}

\textbf{Example play 1.} 
First, we consider an example of Baldwin's fourth-down decision charts.
Figure~\ref{fig:baldwin_ex} illustrates his charts for a fourth-down play in which the Patriots had the ball against the Colts in Week 10 of 2023.
Baldwin views $\go$ as a ``strong'' decision because he estimates that going for it provides a $3.6\%$ gain in win probability over attempting a field goal.

\begin{figure}[p]
    \centering
    \includegraphics[width=0.575\textwidth]{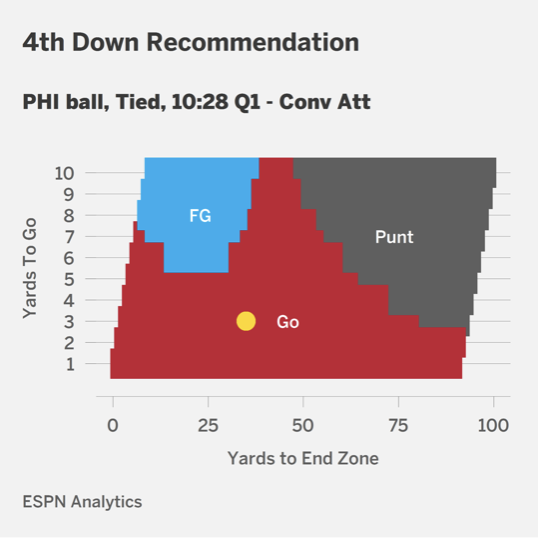}
    \caption{
        Burke's decision boundary chart for example play 2.
        The chart visualizes the estimated optimal decision according to effect size (color) as a function of yards to opponent endzone ($x$-axis) and yards to go ($y$-axis), holding the other game-state variables constant. 
        The yellow dot denotes the actual play's yards to opponent endzone and yards to go.
    }
    \label{fig:burke_ex2_L}
\end{figure}

\textbf{Example play 2.} 
Next, we give an example of Burke's decision chart.
Figure~\ref{fig:burke_ex2_L} illustrates his chart for a play from the 2023 NFC Championship game.
Burke views $\go$ as the right decision because the yellow dot (denoting the actual play's yardline and yards to go) lies squarely in the red region (denoting that $\go$ is the estimated optimal decision) and is far from the decision boundary.

\subsection{Estimating the win probability of a fourth-down decision}\label{sec:fourthd_wp}

An offense has three possible decisions to make on fourth down: it can punt the ball, kick a field goal, or attempt a conversion (colloquially known as ``going for it''), denoted $\punt$, $\fg$, and $\go$, respectively.
In this section we estimate the win probability of a fourth-down decision in $\{\punt, \fg, \go\}$ in terms of first-down win probability (denoted $\wpone$) and decision transition probabilities (punt outcome distribution, field goal success probability, and conversion outcome distribution).
We estimate first-down win probability in Section~\ref{sec:existing_first_down_wp_models} and decision transition probabilities in Section~\ref{sec:dec_transition_probs}.

\textbf{$\punt$ win probability.} 
Suppose the offensive team has possession on fourth down at yardline $y$ and denote the remainder of the game-state by $\bx$.
If the offensive team punts, the opposing team has possession on first down at the subsequent yardline, which we model as a random variable.
Hence the win probability of punting is one minus the opponent's first-down win probability at the next yardline $y'$ and next game-state $\bx'$,
\begin{equation}
 \sum_{y'} (1 - \wpone(\text{yardline } y', \bx')) \cdot \P(\text{yardline after punting is $y'$}|\bx).
\end{equation}
The next game-state $\bx'$ flips the current game-state variables in $\bx$ that are relative to the team with possession (e.g., score differential, team quality metrics, timeouts remaining, etc.) and doesn't alter the other variables (e.g., time remaining).
We re-write this expression in terms of the expectation over the outcome of the punt,
\begin{equation}
\bE_{\text{punt}}[1 - \wpone(\text{yardline } y', \bx') | \bx]. 
\end{equation}
First-down win probability is mostly linear in yardline at most game-states, so for simplicity we instead compute win probability at the expected next yardline after punting,
\begin{equation}
1 - \wpone(\text{yardline } \bE_{\text{punt}}[y'|\bx], \bx').
\label{eqn:v_punt}
\end{equation} 
We model the expected next yardline after punting as a function of yardline and punter quality in Section~\ref{sec:dec_transition_probs}. 

\textbf{$\fg$ win probability.} 
We decompose field goal win probability in terms of field goal success probability and first-down win probability on the subsequent play,
\begin{equation}
 \wp(\text{make }\fg)\cdot \P(\text{make }\fg)  +  \wp(\text{miss }\fg) \cdot (1-\P(\text{make }\fg)).
\label{eqn:v_fg}
\end{equation}
We model the probability of a successful field goal as a function of yardline and kicker quality in Section~\ref{sec:dec_transition_probs}. 
If the kicking team misses the field goal and the spot of the kick is within 20 yards to the opponent's endzone, the opposing team has a first-down possession at $80$ yards to the opponent's endzone.
If the kicking team misses the field goal and the spot of the kick is beyond 20 yards to the opponent's endzone, the opposing team has a first-down possession at the spot of the kick.
The spot of the kick is typically 7 yards behind the line of scrimmage, so in terms of the current yardline $y$, the next yardline relative to the opposing team is $\min\{80,100-(y+7)\}$.
Denoting the remaining game-state variables relative to the opposing team at the next play by $\bx'$, we have
\begin{equation}
 \wp(\text{miss }\fg) = 1-\wpone(\text{yardline } \min\{80,100-(y+7)\}, \bx').
\end{equation}
If the kicking team makes the field goal, it scores three points and the opposing team has a first-down possession after a kickoff.
Denoting the score differential relative to the kicking team by $s$, we have
\begin{equation}
\wp(\text{make }\fg) = 1 - \bE_{\text{kickoff}}[ \wpone(\text{yardline }  y', s' = -s-3, 
\bx')].
\end{equation}
As we did in estimating $\punt$ win probability, for simplicity we instead compute win probability at the expected next yardline after a kickoff,
\begin{equation}
\wp(\text{make }\fg) \approx 1-\wpone(\text{yardline } \bE_{\text{kickoff}}[y'], \ s' = -s-3, \bx').
\end{equation}
The vast majority of kickoffs end in a touchback (yardline $75$),\footnote{
    It is worth noting that the new NFL kickoff rule will change this (see \url{https://operations.nfl.com/the-rules/rules-changes/dynamic-kickoff-rule-explainer/}).
} so for simplicity we instead compute
\begin{equation}
\wp(\text{make }\fg) \approx 1-\wpone(\text{yardline } 75, \ s' = -s-3, \bx').
\end{equation}

\textbf{$\go$ win probability.} 
Suppose the offensive team has possession on fourth down and $z$ yards-to-go at yardline $y$.
If the offensive team goes for it and gains $\Delta \geq z$ yards, then in the next play it either has possession on first down at yardline $y - \Delta$  or scores a touchdown.
Conversely, if the offensive team goes for it and gains $\Delta < z$ yards, then in the next play the opposing team has possession on first down at yardline $100 - (y-\Delta)$.
Hence the expected value of going for it on fourth down is
\begin{equation}
\bE_{\text{go}} \big[ \ind{\Delta \geq z} \cdot \wpone(\text{yardline } y-\Delta) + \ind{\Delta < z} \cdot (1-\wpone(\text{yardline } 100-(y-\Delta))) \big].
\label{eqn:v_go_02}
\end{equation}
This quantity is implicitly a function of the game-state.

This expectation is defined in terms of the conversion outcome conditional density $\allowbreak\P_{\text{go}}(\text{gain $\Delta$ yards}|\bx)$.
This distribution is highly complex, non-normal, and non-symmetric.
For instance, pass plays feature a spike at zero yards gained for incompletions.
The mean yards gained and the tail vary as yardline changes.
There is a spike at the endzone for touchdowns, which increases as the yardline approaches the endzone.
The density differs for each team depending the particularities of a team's play calling and personnel. 
Recent research has gone into estimating this conditional density: \citet{BiroRunDensity} model the conditional density of a run play using a skew-$t$ distribution and \citet{BiroPassDensity} model the conditional density of a pass play using a generalized gamma distribution, which they fit using Markov chain Monte Carlo methods.

For simplicity, we instead switch the order of the expectation and $\wpone$ as we did in estimating $\punt$ and $\fg$ win probability,
\begin{align}
\begin{split}
    &\P(\Delta \geq z|\bx) \cdot \wpone(\text{yardline } y-\bE_{\text{go}}[\Delta|\bx,\Delta \geq z],\bx') \\
+ \ & \P(\Delta < z|\bx) \cdot (1-\wpone(\text{yardline } 100-(y-\bE_{\text{go}}[\Delta|\bx,\Delta < z]),\bx')).
\label{eqn:v_go}
\end{split}
\end{align}
Here, $\bx$ is the game-state on fourth down and $\bx'$ is the game-state on the subsequent play (which, as before, flips the game-state variables that are relative to the team with possession and doesn't alter the other variables).
We model the expected outcome of a successful conversion attempt $\bE_{\text{go}}[\Delta|\bx,\Delta \geq z]$, that of a failed conversion attempt $\bE_{\text{go}}[\Delta|\bx,\Delta < z]$, and conversion probability $\P(\Delta \geq z|\bx)$ in Section~\ref{sec:dec_transition_probs}. 

\subsection{Estimating first-down win probability}\label{sec:existing_first_down_wp_models}

Now, we estimate first-down win probability as a function of game-state.
Win probability estimates arise broadly from one of two classes of models, probabilistic state-space models or statistical models.
On one hand, state-space models simplify the game of football into a series of transitions between game-states.
Transition probabilities are estimated from play-level data and are then propagated into win probability by simulating games.
When implemented correctly, these models are sensible ways to estimate $\wp$. However, they are difficult in practice, as they require: a careful encoding of the convoluted rules of football into a set of states and the actions between those states, careful estimation of transition probabilities, and enough computing power to run enough simulated games to achieve desired granularity.
Each of these are nontrivial.

On the other hand, statistical models are fit entirely from historical data.
Given the results of a set of observed football plays, statistical models fit the relationship between certain game-state variables using data-driven regression or machine learning approaches.
These models are widely used today in football analytics thanks to the accessbility of publicly available play-by-play data (e.g., $\nflfastr$ \citep{nflFastR}) and accessible off-the-shelf machine learning models (e.g., $\xgb$ \citep{xgboost}).
Additionally, due to a perceived abundance of data, flexible machine learning models are viewed as more ``trustworthy'' than previous mathematical models that make more simplifying assumptions. 
For these reasons, the open source win probability models used today in football analytics are statistical / machine learning models, which we focus on in this paper.

\citet{lockNettleton} use a Random Forest ($\rf$) \citep{breimanRF} to estimate win probability from historical data.
The response variable is a binary variable indicating whether the team with possession wins the game.
They model win probability as a function of score differential, game seconds remaining, yards to opponent endzone, down, yards to go, the number of timeouts remaining for each team, pre-game point spread, total points scored, and 
an additional feature to capture the change in impact of score differential over the course of a game,
\begin{equation}
\text{adjusted score} = \frac{\text{score differential}}{\sqrt{1 + \text{game seconds remaining}}}.
\label{eqn:adjusted_score}
\end{equation}
They use a Random Forest of 500 regression trees with parameters $\mtry=2$ and $\nodesize=200$.

\citet{BaldwinWP} uses $\xgb$ \citep{xgboost} to estimate win probability from historical data.
He uses the same binary win/loss response variable as before.
Baldwin models win probability as a function of score differential, game seconds remaining, half seconds remaining, yards to opponent endzone, down, yards to go,
whether the team with possession is at home, whether the team with possession receives the second half kickoff, and the number of timeouts remaining for each team.
He uses two additional features to capture the change in impact of point spread and score differential over the course of a game,
\begin{equation}
\text{spread-time} = (\text{point spread}) \cdot \exp\bigg(-4 \cdot \bigg(1 - \frac{3600}{\text{game seconds remaining}}\bigg)\bigg)
\label{eqn:spread_time}
\end{equation}
and 
\begin{equation}
\text{diff-time-ratio} = (\text{score differential}) \cdot \exp\bigg(-4 \cdot \bigg(1 - \frac{3600}{\text{game seconds remaining}}\bigg)\bigg).
\label{eqn:diff_time_ratio}
\end{equation}
Baldwin includes monotonic constraints for yards to opponent endzone, yards to go, down, score differential, timeouts remaining for each team, spread-time, and diff-time-ratio.
He tunes the $\xgb$ hyper-parameters by minimizing cross validated log-loss \citep{BaldwinWP}.

We also initially considered the generalized additive model ($\gam$) \citep{gam} win probability model from \citet{nflWar} but opted in this paper to focus just on flexible non-parametric models because we expect win probability to be replete with interactions between variables.

The aforementioned models are fit from a historical play-by-play dataset that includes plays from all downs.
Those models fit first-down win probability by including down as a covariate.
We fit an $\xgb$ model just using first-down plays.
This model predicts binary win/loss as a function of score differential, game seconds remaining, pre-game point spread, yards to opponent endzone, receive $2^{nd}$ half kickoff indicator, offensive team's number of timeouts remaining, defensive team's number of timeouts remaining, total score, and 
\begin{equation}
\text{scoreTimeRatio} = \frac{\text{score differential}}{0.01+\text{game seconds remaining}}.
\end{equation}
We include monotonic increasing constraints for score differential, scoreTimeRatio, and offensive timeouts remaining and monotonic decreasing constraints for point spread, yards to opponent endzone, and defensive timeouts remaining.

Now, we compare the out-of-sample predictive performance of various first-down win probability models.
Our full dataset consists of all football plays from 2006 to 2021.
We split our dataset in half by randomly sampling 50\% of the games.
The first-down plays from the first 50\% of these games form the hold-out test set.
All the plays from the other 50\% of these games form the training set.
To tune the $\xgb$ models, we split the training set in half by randomly sampling 50\% of the games from the training set.
The plays from the first 50\% of these games form the $\xgb$ training set, and the remaining plays form the validation set for hyper-parameter tuning.
We then tune our $\xgb$ models in a similar fashion as \citet{BaldwinWP}.
We view the results of our prediction contest in Table~\ref{table:wp_model_selection}.
For reference, we include the predictive performance of two baseline models, a fair coin (always predict 0.5) and a game-level logistic regression model with one linear term for pre-game point spread.
Each model's reduction in error (relative to a fair coin) is the negative percent difference between it's accuracy and the fair coin's.
Our model fit from just first-down plays performs the best out-of-sample. 
Thus, in the remainder of this paper we use our proposed first-down win probability model.

\begin{table}[hbt!]
\centering
\begin{tabular}{lllll} \hline
\vtop{\hbox{\strut model}\hbox{\strut name}} & 
\vtop{\hbox{\strut model}\hbox{\strut type}} &
\vtop{\hbox{\strut plays in}\hbox{\strut training data}} &
\vtop{\hbox{\strut out-of-sample}\hbox{\strut $\logloss$}} &
\vtop{\hbox{\strut reduction }\hbox{\strut in error }} \\ \hline
 Our proposed model & $\xgb$ & first downs & 0.440 & 36.54\% \\
 \citet{lockNettleton} & $\rf$ & all downs & 0.446 & 36.50\% \\
 \citet{BaldwinWP} & $\xgb$ & all downs & 0.476 & 31.27\% \\
 Pre-game point spread only & logistic regression & & 0.609 & 12.10\% \\
 Fair coin & a constant & & 0.693 & 0\% \\ \hline
\end{tabular}
\caption{Predictive performance of first-down win probability models.}
\label{table:wp_model_selection}
\end{table}

\subsection{Estimating fourth-down decision transition probabilities}\label{sec:dec_transition_probs}

Next, we estimate fourth-down decision transition probabilities. 
We model the expected outcome (yardline) of a punt, field goal success probability, conversion success probability, and the expected outcome (yardline) of a successful or a failed conversion attempt.

\begin{figure}[hbt!]
    \centering{}
    \subfloat[\centering \label{fig:punt_model}]{{\includegraphics[width=0.5\textwidth]{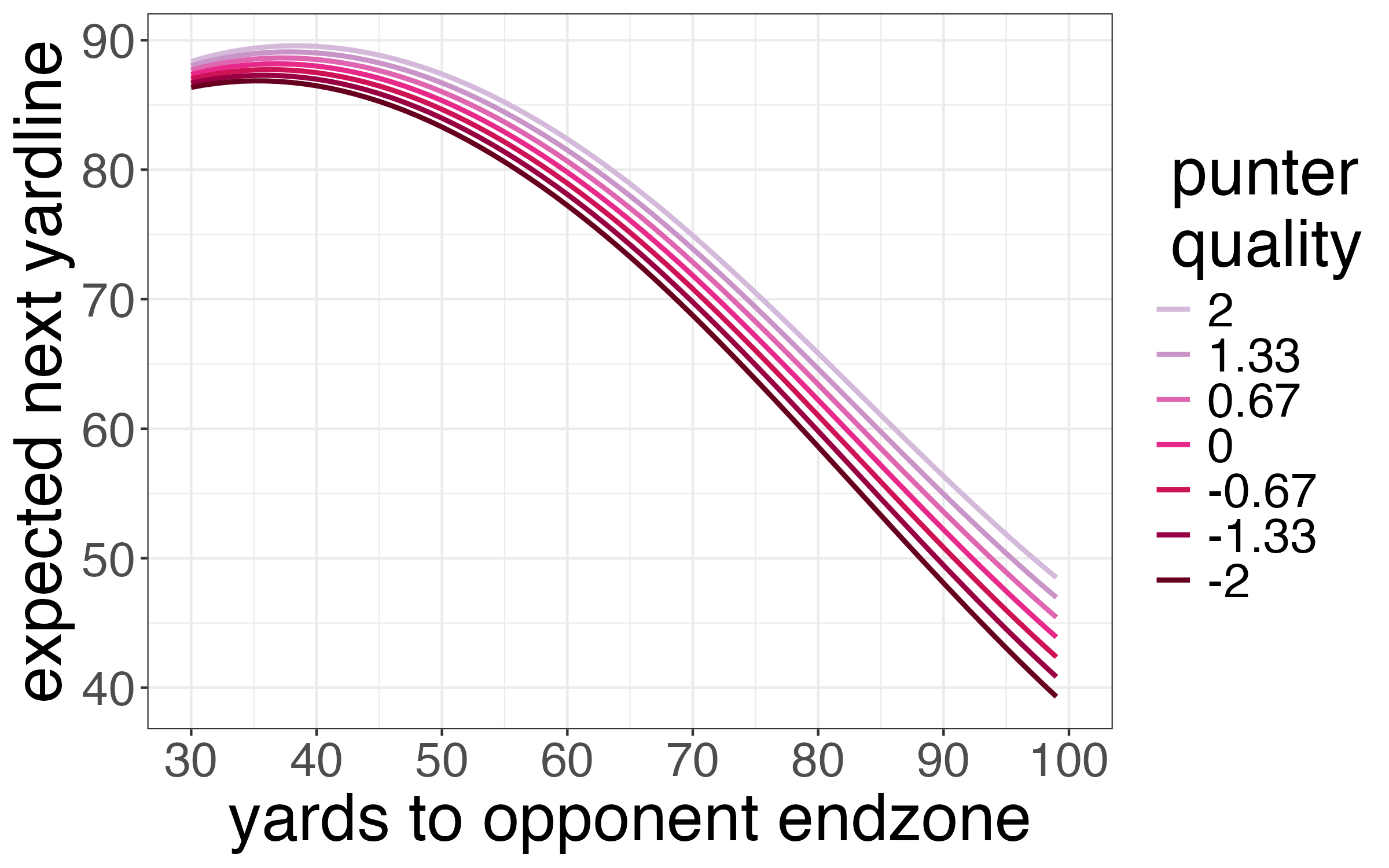} }}
    \subfloat[\centering \label{fig:fgp_model}]{{\includegraphics[width=0.5\textwidth]{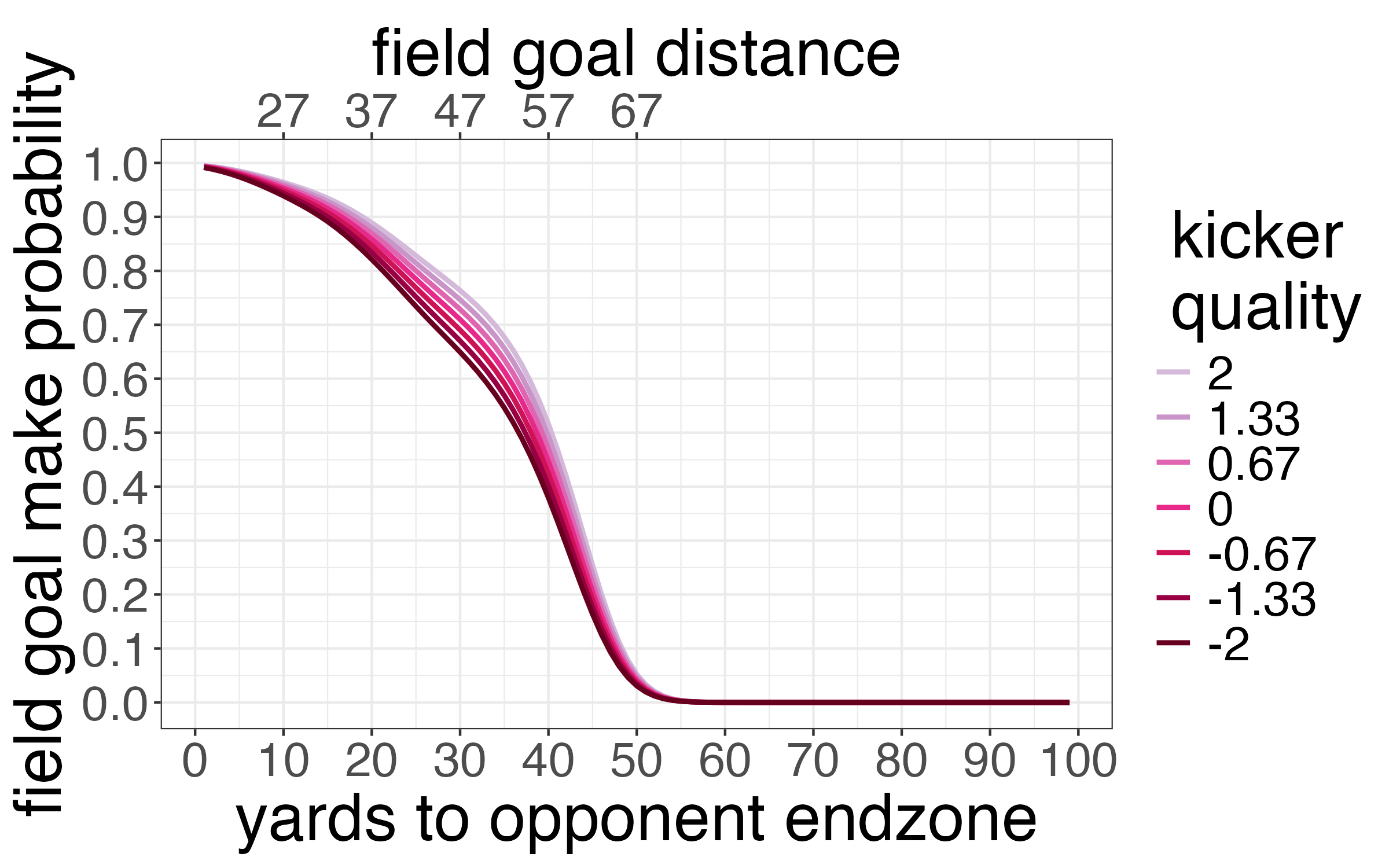} }}
    \caption{
    (a) The expected next yardline after a punt ($y$-axis) according to our model as a function of yardline ($x$-axis) and punter quality (color).  
    (b) The probability of making a field goal ($y$-axis) according to our model as a function of yardline ($x$-axis) and kicker quality (color). 
    }
\end{figure}

\textbf{$\punt$ expected outcome model.} 
We use linear regression to model the expected next yardline after a punt as a function of yardline and punter quality (denoted $\pq$, whose specification is detailed in Appendix~\ref{app:player_quality}),
\begin{align}
\begin{split}
    \bE_{\text{punt}}[\text{next yardline}] &= \ 
    \vec{\alpha} \cdot \text{B-spline}\big(\text{yardline}, \ \text{df}=4\big) + \ \beta_1\cdot\pq + \beta_2\cdot\pq\cdot\text{yardline}.
\label{eqn:punt_model}
\end{split}   
\end{align}
We use a B-spline to capture a nonlinear trend in yardline \citep{splines}.
The model is trained on a dataset of $36,493$ punts from 2006 to 2021, all beyond the 30 yardline.
We visualize this model in Figure~\ref{fig:punt_model}.
The above expectation is more properly defined in terms of the punt outcome conditional density $\allowbreak\P_{\text{punt}}(\text{next yardline}|\bx)$.
But, that distribution is difficult to model (recall our discussion in Section~\ref{sec:fourthd_wp} of modeling the conditional density of yards gained).
A more sophisticated analysis would use that conditional density.


\textbf{$\fg$ success probability model.} 
We use logistic regression to model the probability that a kicker makes a field goal as a function of yardline and kicker quality (denoted $\kq$, whose specification is detailed in Appendix~\ref{app:player_quality}),
\begin{align}
\begin{split}
    \log\bigg( \frac{ \P(\text{make }\fg) }{1 - \P(\text{make }\fg) } \bigg) = \
    \vec{\alpha} \cdot \text{B-spline}\big(\text{yardline}, \ \text{df}=5\big) + \beta\cdot\kq.
\label{eqn:fgp_model}
\end{split}   
\end{align}
Fitting this model on our dataset of $15,472$ observed field goals from 2006 to 2021 yields nontrivial probability predictions for extremely long field goals that have never before been made (e.g., nontrivial probability for a 73-yard field goal from the 55 yards to the opponent's endzone).
To shrink these field goal probability predictions to zero, we impute 500 synthetic missed field goals, randomly distributed from the 51 to 99 yards to the opponent's endzone, into our dataset. 
We visualize this model in Figure~\ref{fig:fgp_model}.

\textbf{$\go$ conversion probability model.} 
Due to a small sample size of fourth-down conversion attempts, some existing fourth-down conversion probability models use third down as a proxy for fourth down (e.g., \citet{Romer06dofirms}), as they are also high-pressure situations in which the offensive team attempts to obtain a first down.
But, there may be a fundamental difference in conversion probability between third and fourth-down plays, perhaps due to psychological reasons or a differing distribution of play calls.
In our model selection process, we train some models on a dataset consisting entirely of fourth-down plays and other models on a dataset consisting of third and fourth-down plays, testing models on a random 50\% of fourth-down plays.
The parameters of our best conversion probability model borrow strength from third-down plays.

\begin{figure}[hbt!]
    \centering{}
    {\includegraphics[width=0.8\textwidth]{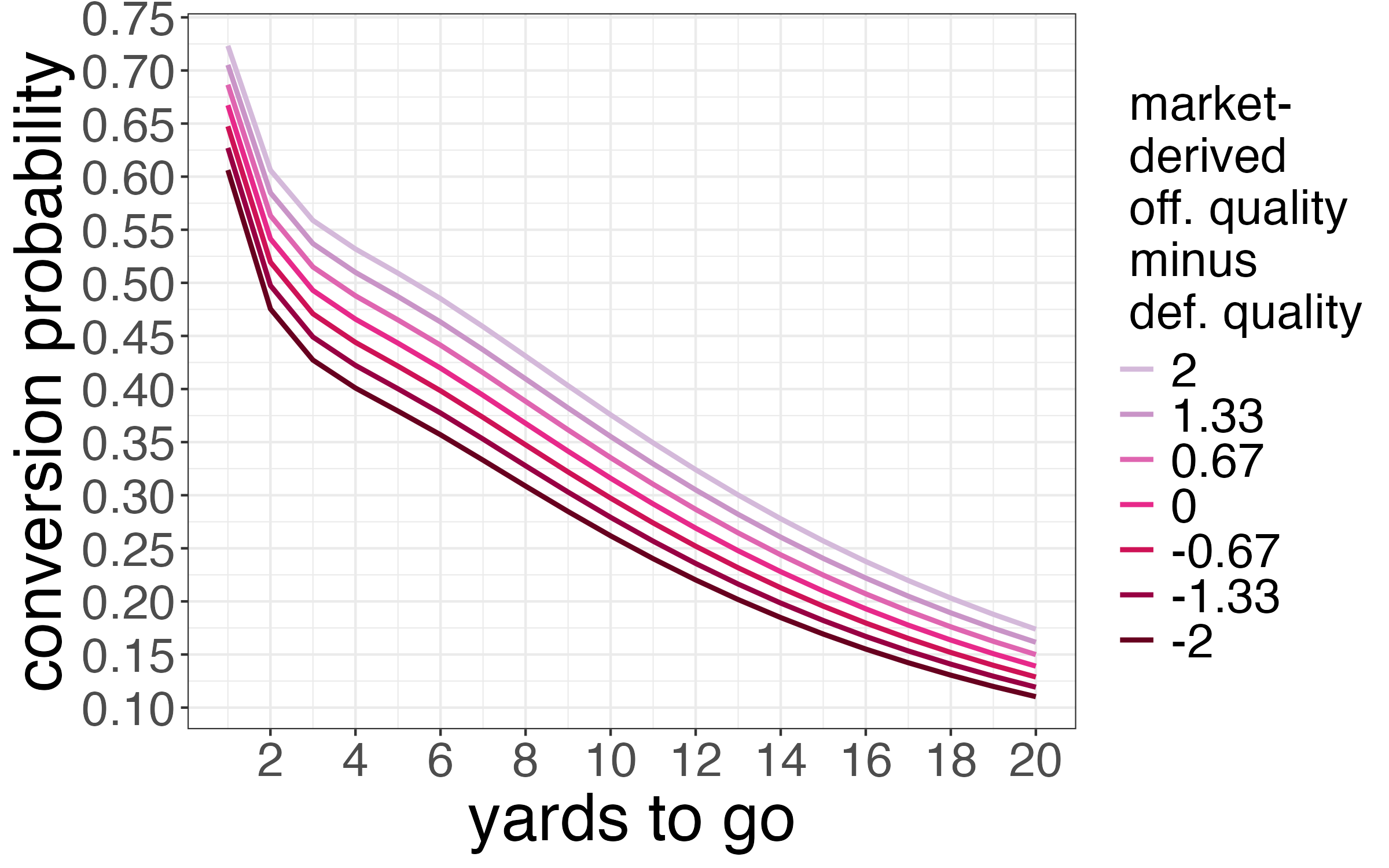} }
    \caption{
    Fourth-down conversion probability ($y$-axis) according to our model as a function of yards to go ($x$-axis) and the difference in offensive and defensive quality (color). 
    }
    \label{fig:conv_model_tq}
\end{figure}

We use logistic regression to model fourth-down conversion probability as a function of yards to go, down (third vs. fourth down), and a market-derived measure of how much better the offensive team's offensive quality is than the defensive team's defensive quality (denoted $\Delta\tq$, or the difference in team quality, whose specification is detailed in Appendix~\ref{app:player_quality}).
Formally, we model
\begin{align}
    \log\bigg( \frac{ \P({\text{convert}}) }{1 - \P({\text{convert}}) } \bigg) =& \ 
    \vec{\alpha_1} \cdot \ind{\text{fourth down}} \cdot \text{B-spline}\big(\log(\text{yards to go}+1), \ \text{df}=4\big) \\
    +& \ \vec{\alpha_2} \cdot \ind{\text{third down}} \cdot \text{B-spline}\big(\log(\text{yards to go}+1), \ \text{df}=4\big) \\
    +& \ \beta_0 + \beta_1 \cdot \Delta\tq.    
\label{eqn:conv_model}
\end{align}
We visualize this model in Figure~\ref{fig:conv_model_tq}.

\textbf{$\go$ expected outcome models.} 
We use linear regression to model the expected outcome (yards gained $\Delta$) of a conversion attempt with $z$ yards to go given that it is successful by
\begin{align}
\begin{split}
    \bE_{\text{go}}[\Delta|\bx,\Delta \geq z] =& \ 
    \vec{\alpha_1} \cdot \ind{\text{fourth down}} \cdot \text{B-spline}\big(\log(\text{yards to go}), \ \text{df}=4\big) \\
    +& \ \vec{\alpha_2} \cdot \ind{\text{third down}} \cdot \text{B-spline}\big(\log(\text{yards to go}), \ \text{df}=4\big) \\
    +& \ \vec{\alpha_3} \cdot \ind{\text{yards to go}=1} \cdot \text{B-spline}\big(\text{yardline}, \ \text{df}=3\big) \\
    +& \ \vec{\alpha_4} \cdot \ind{\text{yards to go} \neq 1} \cdot \text{B-spline}\big(\text{yardline}, \ \text{df}=4\big) \\
    +& \ \beta_0 + \beta_1 \cdot \Delta\tq.    
\label{eqn:conv_expected_outcome_success_model}
\end{split}   
\end{align}
Similarly, we model the expected outcome of a conversion attempt  with $z$ yards to go given that it is unsuccessful by 
\begin{align}
\begin{split}
    \bE_{\text{go}}[\Delta|\bx,\Delta < z] =& \ 
    \alpha_1 \cdot \ind{\text{fourth down}} \cdot \log(\text{yards to go}+1) \\
    +& \ \alpha_2 \cdot \ind{\text{third down}} \cdot \log(\text{yards to go}+1) \\
    +& \ \beta_0 + \beta_1 \cdot \Delta\tq.    
\label{eqn:conv_expected_outcome_failure_model}
\end{split}   
\end{align}
These two expectations are more properly defined in terms of the conversion outcome conditional density $\allowbreak\P_{\text{go}}(\text{gain $\Delta$ yards}|\bx)$.
But, as discussed in Section~\ref{sec:fourthd_wp}, that distribution is highly complex, non-normal, and non-symmetric, making it difficult to model.
A more sophisticated analysis would use that conditional density.


\section{Fourth-down decision making under uncertainty}\label{sec:uncertainty_sec}

The standard approach, recommending the decision that maximizes estimated win probability, disregards the uncertainty that arises from estimating win probability from a finite dataset.
Thus, in this section, we modify the fourth-down decision procedure to account for uncertainty. 

Analysts may be inclined to trust estimates from win probability models because they are fit from an ostensibly massive dataset: there are $600,825$ plays and $229,635$ first-down plays in our play-by-play dataset.
Nonetheless, several characteristics of this dataset reveal just how difficult it is to estimate win probability. 
First, the binary win/loss outcome variable is coarse.
Since win probability is a continuous quantity in the interval $[0,1]$, to capture it with fine granularity from raw zeros and ones requires a massive amount of data.
Estimating win probability as a function of game-state, with such a large set of possible game-states, dramatically increases the difficulty of the task.
Win probability consists of complex nonlinear and interacting relationships between game-state variables, not simple additive relationships, which makes it even more difficult.

Moreover, the binary win/loss response variable is clustered: all plays from the same game share the same win/loss outcome.
This dependence structure reduces the effective sample size, which is somewhere between the number of first-down plays ($229,635$) and the number of non-tied games ($4,101$) in our dataset.
According to the simulation study in \citet{brillSimWP}, we effectively have half as much data as we think.
In other words, a win probability estimator fit from a dataset consisting of half as many plays, but where each play has an independent outcome, has the same accuracy as an estimator fit from a clustered dataset like our historical play-by-play dataset. 

Thus, there are likely not enough data to experience the full variability of the nonlinear and interacting variables of score differential, time remaining, point spread, yards to opponent endzone, yards to go, timeouts, etc.
In fitting win probability models, we are in a limited-data context.
Therefore, it is imperative to incorporate uncertainty quantification into the fourth-down decision procedure.

Since we estimate win probability using machine learning models, we need a non-parametric method to quantify uncertainty in these estimates.
Bootstrapping is a natural choice to capture such uncertainty.
The bootstrapping process begins with generating $B$ bootstrapped datasets from our original dataset.
To do so, we use a randomized cluster bootstrap: re-sample games uniformly with replacement and within each game re-sample drives uniformly with replacement.
According to the simulation study in \citet{brillSimWP}, although the randomized cluster bootstrap achieves higher coverage than the standard i.i.d. bootstrap (re-sample plays uniformly with replacement), it produces undercovered intervals that are too narrow.
Thus, we also tune our randomized cluster bootstrap using a fractional bootstrap as recommended in \citet{brillSimWP}.
We choose an appropriate value for $B$ ($B=101$) in Appendix~\ref{app:choosing_B}. 

Then, to each bootstrapped dataset, which represents a re-draw of the recent history of football, we fit a full fourth-down decision model (described in Section~\ref{sec:wp}).
At each game-state, each model yields an estimated optimal decision.
To quantify uncertainty in the estimated optimal decision, we bag the decision itself.
Specifically, we consider the \textit{bootstrap percentage} ($\bootp$), or the percentage of bootstrapped models that report each decision to be optimal.
Formally, at game-state $\bx$ we have the original estimated optimal decision $d(\bx) \in \{\go,\fg,\punt\}$ (the point estimate) and $B$ bootstrapped estimated optimal decisions $d_{1}(\bx),...,d_{B}(\bx)$.
Then
\begin{equation}
    \bootp(\bx) = 100\% \cdot \frac{1}{B}\sum_{b=1}^{B} \ind{ d_{b}(\bx) = d(\bx) }.
\end{equation}

Bootstrap percentage quantifies how confident we are that a recommendation has higher win probability than the other decisions.
A high $\bootp$ ($\approx 100\%$) reflects high confidence: the estimated optimal decision remains the same across the vast majority of re-draws of the training set.
Conversely, a low $\bootp$ ($\approx 50\%$) reflects low confidence: the estimated optimal decision differs substantially across different re-draws of the training set.
A low $\bootp$ is $\approx 50\%$ or lower because a fourth-down decision almost always boils down to picking between $\go$ and one of the kicks, as it is clearly sub-optimal to punt near the opponent's endzone and to attempt a field goal far from the opponent's endzone.

Regardless of the effect size, when $\bootp$ is low we cannot trust the point estimate: the effect size at that game-state is too dependent on the random idiosyncrasies of its particular training set.
In other words, the edge detected by the point estimate is more due to noise than signal.

We also consider confidence intervals on the effect size $g(\bx)$, the gain in win probability by making a fourth-down decision at game-state $\bx$.
Given the original estimated effect size $\hat{g}(\bx)$ and the $B$ bootstrapped effect sizes $\hat{g}_{1}(\bx),...,\hat{g}_{B}(\bx)$, sorted from least to greatest, we construct a confidence interval on the effect size using the quantiles.
For $B=101$, a $90\%$ confidence interval on the effect size is $[\hat{g}_{6}(\bx), \ \hat{g}_{96}(\bx)]$.
This confidence interval represents the following:
assume that the win/loss outcomes across football history are generated by some underlying win probability function.
If we re-simulated the history of football $B$ times from that $\wp$ function, keeping the pre-game conditions of each game the same, the true win probability is expected to lie in the $90\%$ confidence interval $0.9 \cdot B$ times.

We find it easier for bootstrap percentage to drive decision making than confidence intervals because the former is defined by one number whereas the latter is defined by two numbers.
For example, consider a play in which the estimated gain in win probability for $\go$ is $1\%$.
If our confidence interval for this gain is $[-3\%, 5\%]$, it is not clear how strongly we should recommend $\go$. 
Similarly, it is difficult to compare two confidence intervals of varying lengths.
For example, it is not clear whether our recommendation for a confidence interval of $[-3\%, 5\%]$ should be weaker than that for a confidence interval of $[-1\%, 3\%]$.
Also, our primary interest is to quantify uncertainty in the decision itself \citep{wynerBootDecision}, which is not as granular as quantifying uncertainty in the effect size.

\section{Results}\label{sec:fourth_down_dec}

In the previous section we modified the decision procedure to account for uncertainty.
In Section~\ref{sec:example_plays} we illustrate this amended procedure using example plays.
Then, in Section~\ref{sec:overconfidence} we consider the extent to which the traditional decision procedure was overconfident in its fourth-down recommendations.

\subsection{Example plays}\label{sec:example_plays}

Now, we discuss example plays.
To compare decision recommendations to the decisions that actual football coaches tend to make, we model the probability that a coach chooses a decision in $\{\go,\fg,\punt\}$ as a function of game-state.
We detail the specification of this \textit{baseline coach model} in Appendix~\ref{app:baseline_coach_model} and include the model's predictions in our decision figures. 

\begin{figure}[hbt!]
    \centering
    \subfloat[\centering \label{fig:our_decisions_ex_1A}]{ 
        \includegraphics[width=1\textwidth]{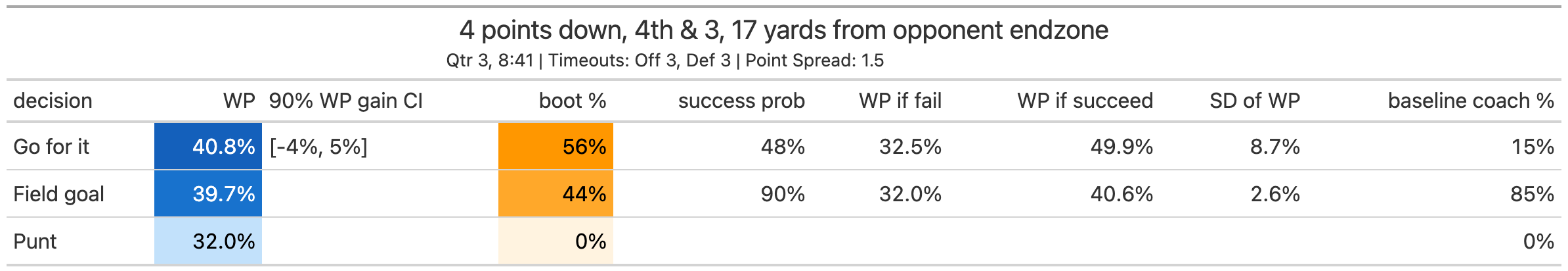}
    } \\
    \subfloat[\centering \label{fig:our_decisions_ex_1B}]{ 
        \includegraphics[width=0.5\textwidth]{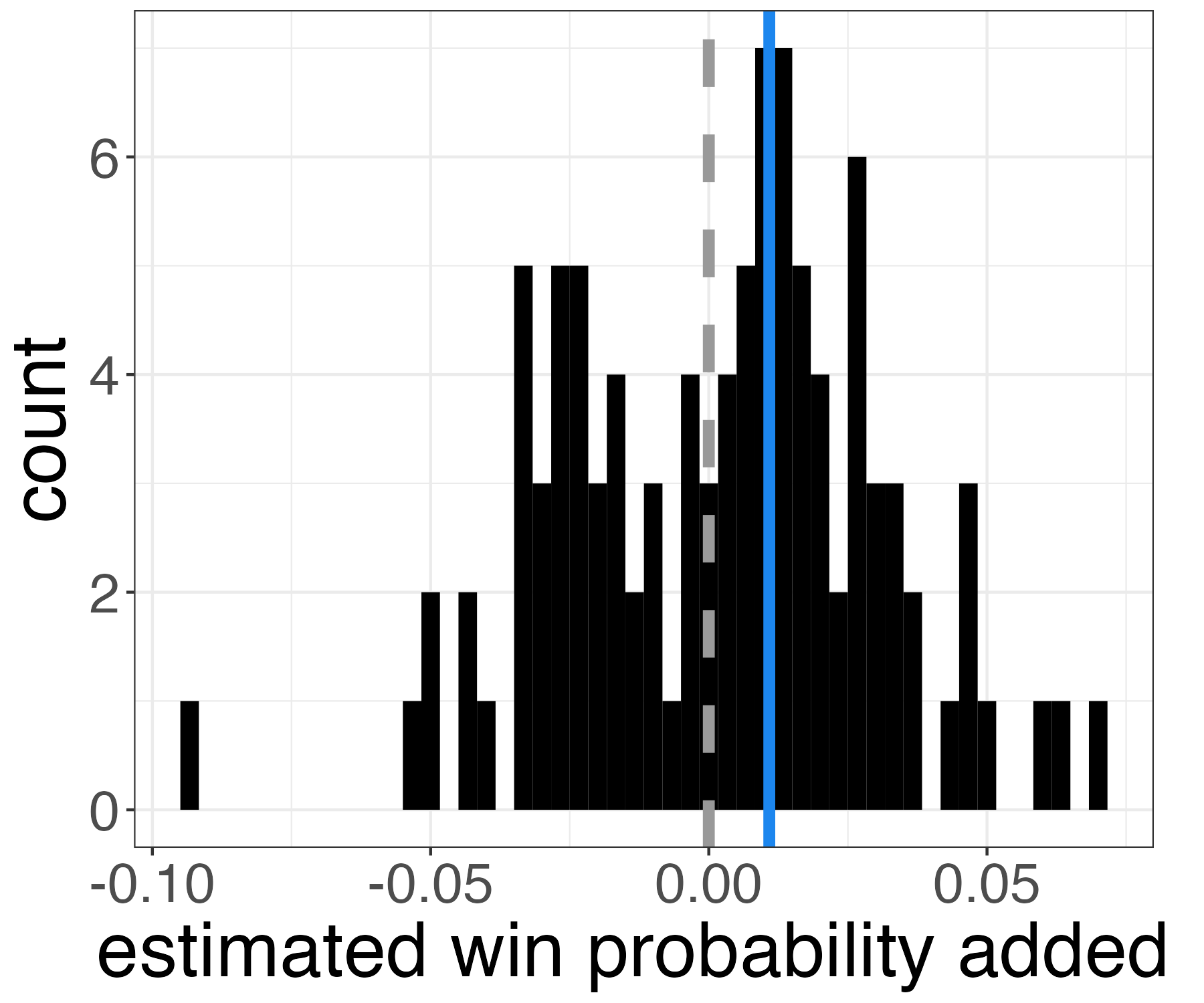}
    }
    \caption{
        Decision charts for example play 1.
        Figure (a) shows the models' outputs, including the point estimate (blue column), $\wp$ gain confidence interval, $\bootp$ (orange column), and other metrics.
        Figure (b) is the histogram of bootstrapped win probability gain estimates. The gray dashed line is zero and the blue solid line is the point estimate.
    }
    \label{fig:our_decisions_ex_1}
\end{figure}

\textbf{Example play 1.} 
First, recall example play 1 from Section~\ref{sec:trad_examples} in which the Patriots had the ball against the Colts in Week 10 of 2023.
In Figure~\ref{fig:our_decisions_ex_1} we knit uncertainty quantification into decision making. 
Although the point estimate (the blue column in Figure~\ref{fig:our_decisions_ex_1A}) suggests that $\go$ provides a $1.1\%$ gain in win probability over $\fg$, $44\%$ of the bootstrapped models view $\fg$ as better than $\go$ (the orange column in Figure~\ref{fig:our_decisions_ex_1A}), reflecting substantial uncertainty in the optimal fourth-down decision.
Also, the $90\%$ confidence interval is $[-4\%, 5\%]$, and the histogram of bootstrapped win probability gain estimates in Figure~\ref{fig:our_decisions_ex_1B} is centered near zero and has a large spread.
These suggest that $\go$ could either be a good or a bad decision. 


\begin{figure}[hbt!]
    \centering
    \subfloat[\centering \label{fig:burke_ex2_M}]{ 
        \includegraphics[width=0.5\textwidth]{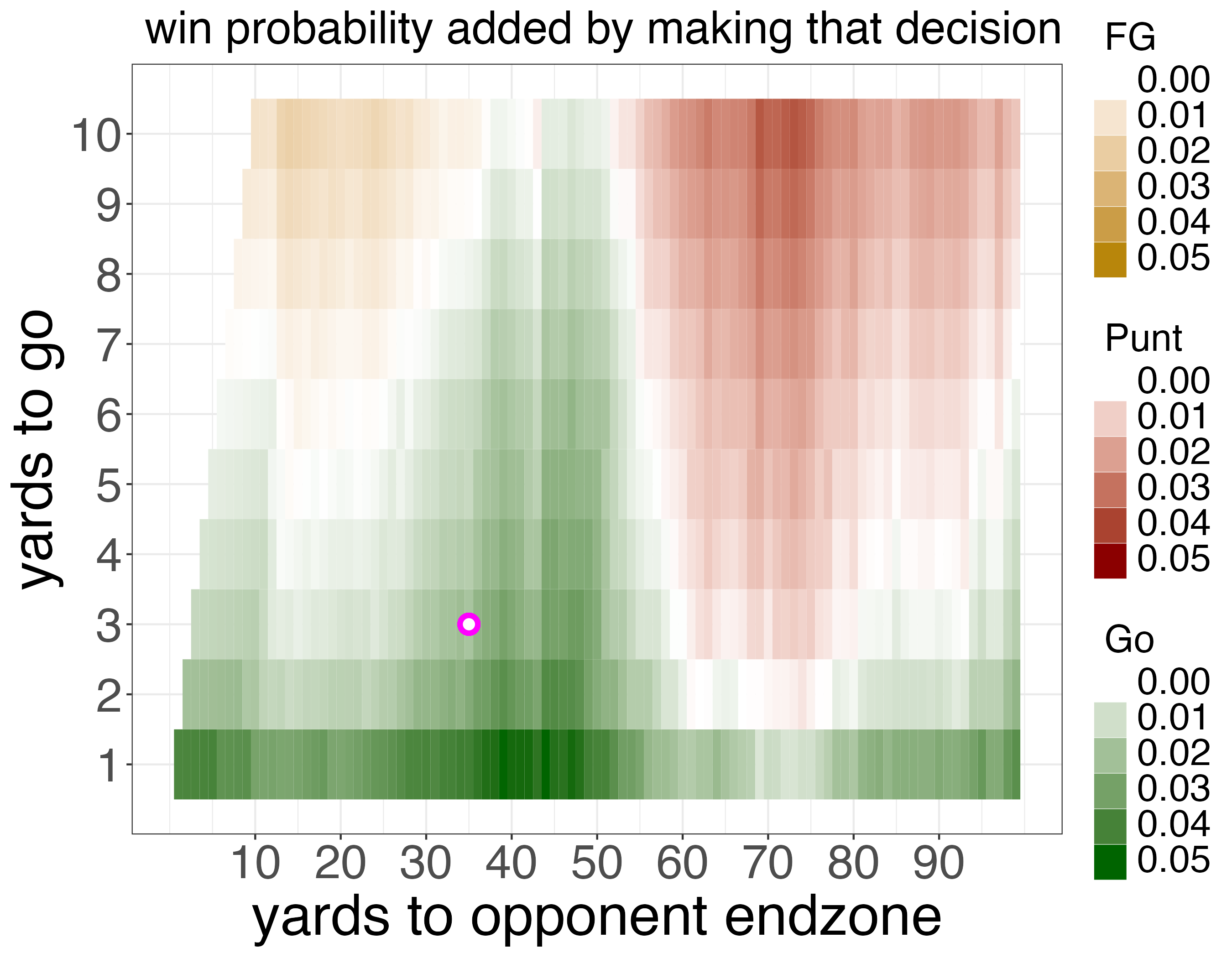}
    }
    \subfloat[\centering \label{fig:burke_ex2_R}]{ 
        \includegraphics[width=0.5\textwidth]{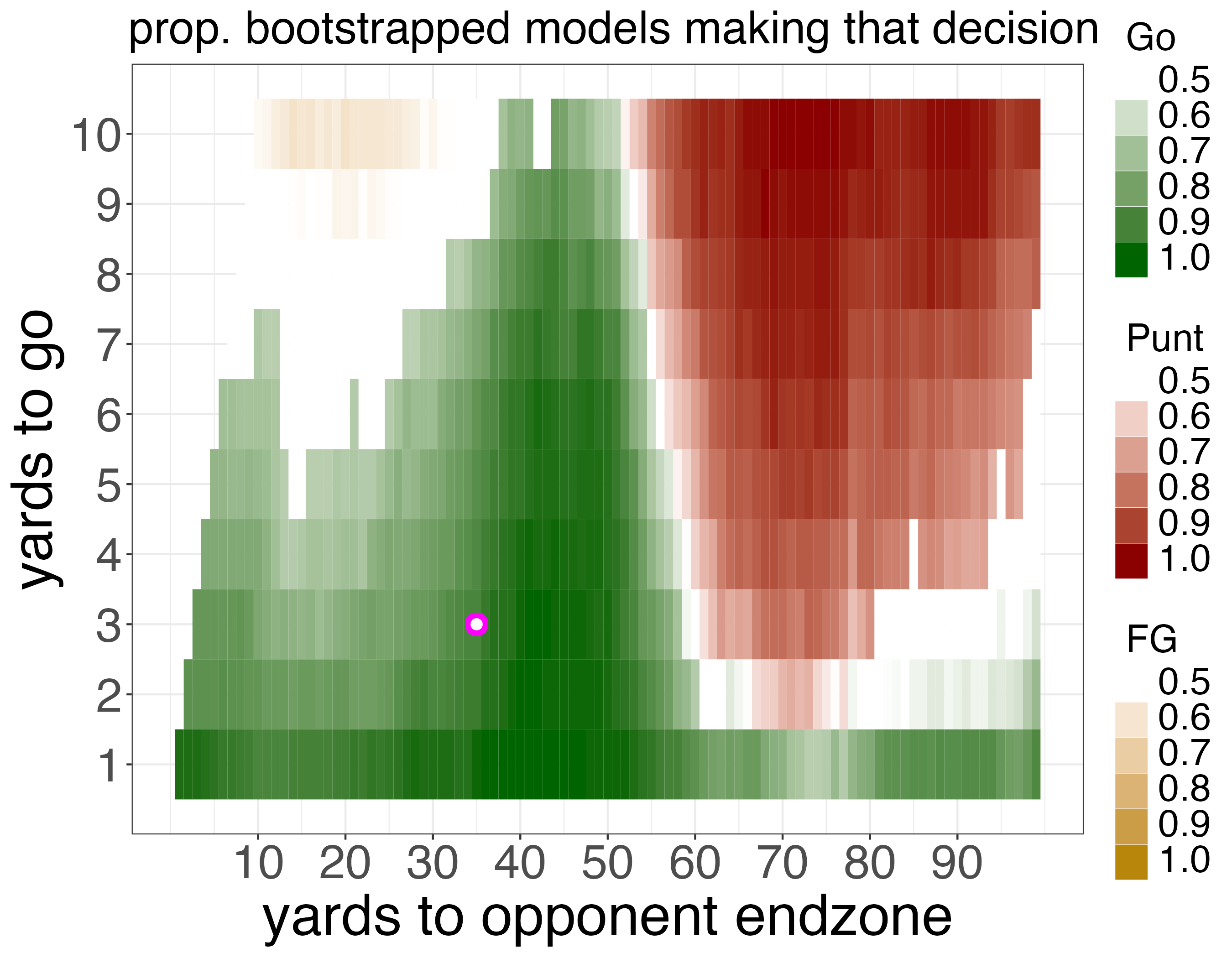}
    }
    \caption{
    Decision boundary charts for example play 2.
    Figure (a) visualizes the estimated optimal decision (color) according to effect size (color intensity, where darker colors indicate larger values) as a function of yards to opponent endzone ($x$-axis) and yards to go ($y$-axis), holding the other game-state variables constant.
    The pink dot represents the actual play's yards to opponent endzone and yards to go.
    Figure (b) is similar except color intensity reflects bootstrap percentage.
    }
    \label{fig:burke_ex2}
\end{figure}

\textbf{Example play 2.} 
Next, recall example play 2 from Section~\ref{sec:trad_examples} in which the Eagles had the ball against the 49ers in the 2023 NFC Championship game.
We visualize fourth-down decision recommendations in Figure~\ref{fig:burke_ex2}.
The pink dot in Figure~\ref{fig:burke_ex2_M} lies in a moderately dark green region, indicating a moderate estimated gain in win probability by going for it.
Being far from the decision boundary, however, does not imply it the best decision with certainty.
Also, it is not obvious how far from the boundary is ``far enough.''
Hence, in Figure~\ref{fig:burke_ex2_R} we provide an additional chart that illustrates uncertainty in the estimated optimal decision.
Here, the color intensity indicates the proportion of bootstrapped models that make the estimated optimal decision.
Much of the figure consists of dark colors, specifically dark green in the bottom-left and middle and dark red in the top-right.
For those combinations of yardline and yards to go, we are confident in the estimated optimal decision.
For example play 2 (the pink dot), $\bootp$ is high (dark green), so we are confident that $\go$ is the best decision.
Other combinations of yardline and yards to go feature lighter colors, reflecting higher uncertainty.
For those game-states, we are less confident in the estimated optimal decision.

We include more examples of this fourth-down decision procedure in Appendix~\ref{app:additional_ex_plays}.
We also include an interactive Shiny App that visualizes the decision procedure for any game-state on Github.\footnote{
    To run the Shiny app, download the Github repository at \url{https://github.com/snoopryan123/fourth_down} and run \texttt{3\_shiny/app.R}.
} 

\subsection{Quantifying overconfidence in win probability point estimates}\label{sec:overconfidence}

\begin{figure}[hbt!]
    \centering
    \includegraphics[width=0.8\textwidth]{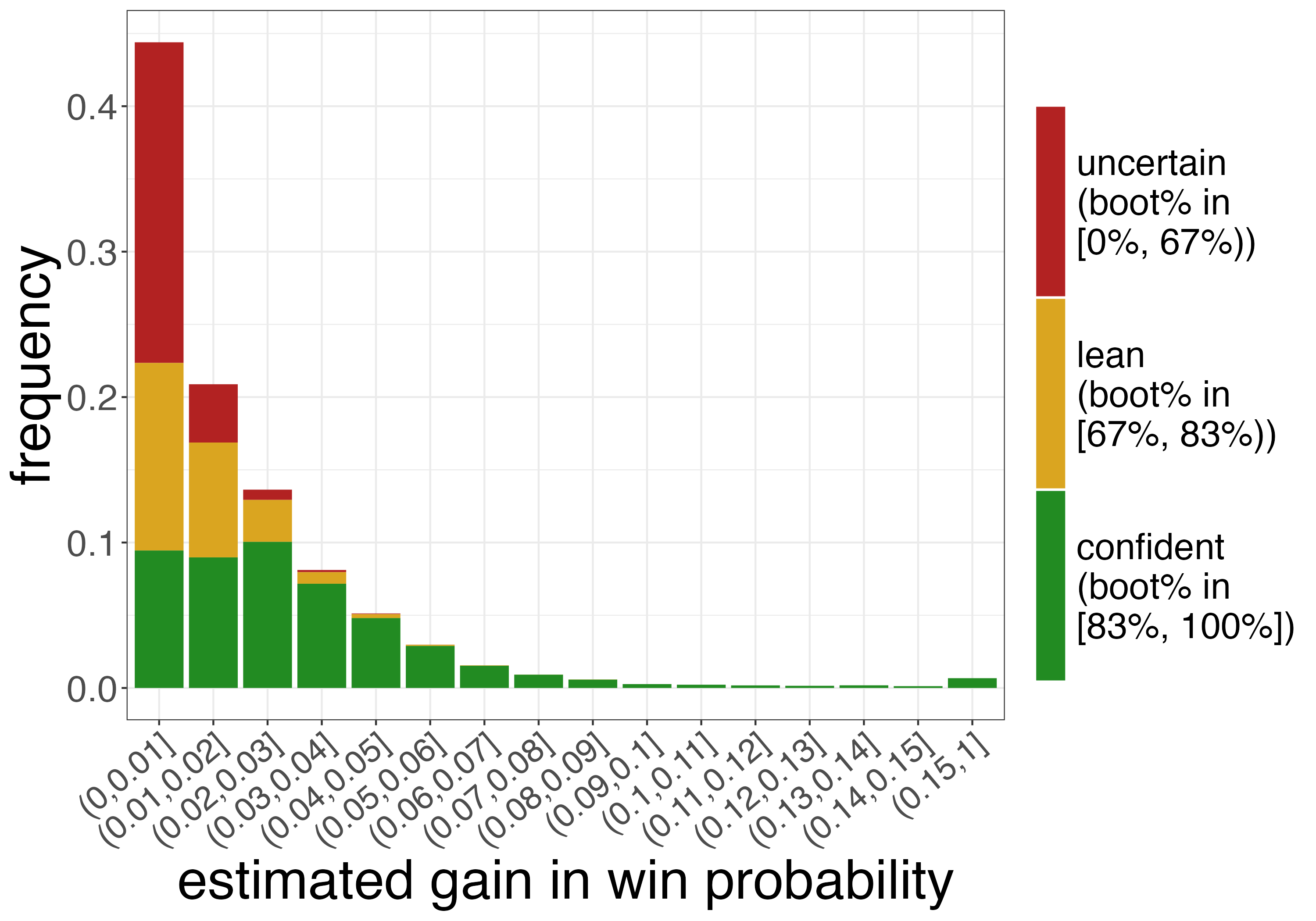}
    \caption{
    The distribution of effect size, and uncertainty given effect size, across all fourth-down plays from 2018-2022.
    The height of a bar reflects the proportion of plays that lie in its associated effect size bin ($x$-axis).
    The color distribution within a bar reflects the distribution of decision confidence (confident in green, lean in yellow, and uncertain in red) among those plays.
    }
    \label{fig:plot_overconfidence}
\end{figure}

In Figure~\ref{fig:plot_overconfidence} we visualize the distributions of effect size and uncertainty across all fourth-down plays from 2018-2022.
Effect size is the estimated gain in win probability and uncertainty is proportional to bootstrap percentage.
To facilitate easier communication to a non-technical football coach, we bin decision confidence levels.
We partition $[50\%, 100\%]$ into three equally spaced buckets and add $[0\%, 50\%)$ to the lowest bucket: confident ($\bootp \in [83\%, 100\%]$), lean ($\bootp \in [67\%, 83\%)$), and uncertain ($\bootp \in [0\%, 67\%)$).

We are confident in the vast majority of decisions that have an effect size above $4\%$.
Most decisions with an effect size under $1\%$ are uncertain, and decisions with an effect size between $1\%$ and $4\%$ are a healthy split between confident, lean, and uncertain. 
Many plays ($44\%$) have an effect size under $1\%$ and most plays ($87\%$) have an effect size under $4\%$.
So, many fourth-down plays feature an effect size that is subject to considerable uncertainty.
In particular, we are confident in just $48\%$ of all fourth-down decisions from 2018 to 2022 and $27\%$ of them are uncertain.
This analysis reflects substantial overconfidence in win probability point estimates; far fewer fourth-down decisions are as obvious as traditional analyses suggest. 

Interestingly, we are confident in about $20\%$ of decisions with effect size under $1\%$ and in about $50\%$ of decisions with effect size between $1\%$ and $2\%$.
Although decision confidence is correlated with estimated gain in win probability, they are fundamentally different: we can be confident in plays that provide small edges.
Over the course of a season, these edges can accumulate into a large advantage that coaches should take advantage of.

\begin{figure}[hbt!]
    \centering
    \includegraphics[width=0.8\textwidth]{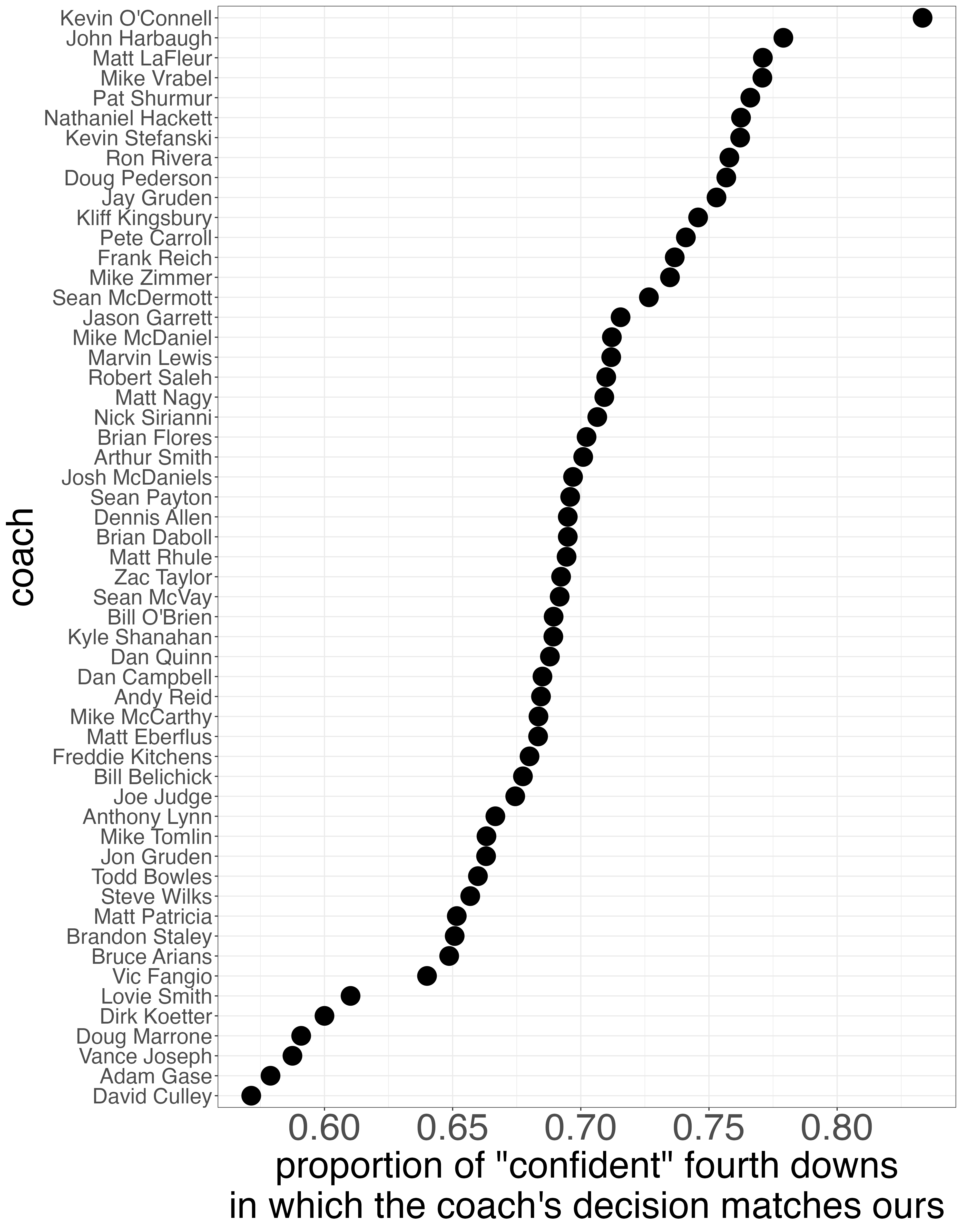}
    \caption{
    NFL coach ($y$-axis) versus the proportion of confident fourth-down decisions he made in agreement with our model from 2018 to 2022 ($x$-axis).
    }
    \label{fig:coach_eval}
\end{figure}

Furthermore, a football analyst shouldn't penalize a coach for making a decision that has high uncertainty regardless of the effect size because the estimated edge may be due to noise.
Accordingly, we should evaluate coaches only on plays for which we are confident (and perhaps also plays in which we lean towards a decision).
In Figure~\ref{fig:coach_eval} we rank coaches by the proportion of fourth-down decisions they made in accordance with our model from 2018 to 2022 among fourth-down plays we are confident in.
The top of this list has coaches like John Harbaugh from ``analytics bent'' organizations but also coaches who are considered more traditional on fourth down such as Ron Rivera and Pat Shurmur.
While ``analytics bent'' coaches tend to follow fourth-down recommendations from win probability point estimates, which include recommendations on confident plays, some of the more traditional coaches seem to have a good feel for obvious fourth downs.

Despite overconfidence in win probability point estimates, analysts have been largely correct that NFL coaches do not go for it enough on fourth down.
Across all fourth-down plays from 2018-2022 that we are confident in, the coach made the right decision for $91\%$ of plays where they should have kicked but just $49\%$ of plays where they should have gone for it. 
Play calling in the NFL is still far too conservative:  coaches consistently make wrong decisions, particularly when they should go for it.

\section{Discussion}\label{sec:discussion}

Win probability models are commonly used across sports analytics.
Fourth-down decision making is a prominent example.
There has been little reflection on the high-variance nature of win probability estimators and no substantial attempt to quantify uncertainty in these models.
In this study, we argue it is imperative to knit uncertainty quantification into win probability-based decision making.
Using fourth-downs as our case study, we in particular recommend a fourth-down decision when we are confident it has higher win probability than all other decisions.
If we are not confident in a recommendation because it is subject to substantial uncertainty, we should not penalize a coach who doesn't follow it.
Similarly, we do not think it is right for a football analyst to recommend an uncertain fourth-down recommendation, regardless of the effect size.
Despite analysts' overconfidence in win probability point estimates, after accounting for uncertainty we still find that NFL coaches are too conservative: they don't go for it enough on fourth down.
Additionally, as teams across the NFL have increased their propensity to go for it on fourth down, the success rate hasn’t declined.\footnote{
    \url{https://medium.com/@dacr444/analyzing-4th-down-attempts-over-the-last-24-seasons-8c0f99c538cb}).
}

Our analysis is not without limitations and there are many avenues for future work.
To begin, in this study, we used bootstrapping to quantify uncertainty in win probability estimates.
Although our bootstrap procedure produces wide confidence intervals, it underestimates uncertainty since it quantifies sampling uncertainty but not model uncertainty.
The former is uncertainty resulting from fitting a model on a finite dataset (``variance'') and the latter is uncertainty caused by our model being wrong or biased (``bias'').
Our models are subject to several sources of model uncertainty.
We made several simplyfing assumptions in modeling fourth-down win probability, such as swapping the expectation with first-down win probability.
Furthermore, ``true'' win probability is a function of unobserved confounders.
For example, the yards to go variable would ideally be derived from tracking data and measured in inches, but the publicly available version is integer-valued.
\citet{lopez2020} shows that, particularly at short distances (i.e., comparing $4^{th}$ down and inches to $4^{th}$ and 1), a more fine-grained yards to go variable can produce substantially different forecasts.
Another key unobserved confounder is the play call that's being considered for a potential conversion attempt.
For instance, on $4^{th}$ and 1, the Eagles ``tush push'' play call has a much higher success rate than the league-wide base rate.\footnote{
    \url{https://www.espn.com/nfl/story/_/id/41234531/philadelphia-eagles-tush-push-jason-kelce}
}
The play call can make or break a conversion attempt, and not adjusting for it introduces additional uncertainty into the analysis. 
Unfortunately, it is not publicly available.
A more elaborate analysis would capture model uncertainty. 

In this study, we found that statistical win probability models produce uncertain estimates at many game-states.
In future work we suggest exploring probabilistic state-space models to estimate win probability.
As discussed in Section~\ref{sec:existing_first_down_wp_models}, probabilistic models simplify the game of football into a series of transitions between game-states.
Transition probabilities are estimated from play-level data and win probability is calculated by simulating games.
The effective sample size ($\ESS$) of transition probability models is the number of plays because they are fit from independent play-level observations.
Some analysts in industry have created proprietary probabilistic models, which they believe are more accurate than statistical models due to the larger $\ESS$.
But, state-space models have higher bias than statistical models.
Whereas both classes of models have similar levels of hidden bias (due to not adjusting for unobserved confounders), state-space models have more model bias as they make stronger simplifying assumptions.
See Appendix~\ref{app:stateSpaceModelBias} for a further discussion of the difficulty of formulating probabilistic state-space win probability models.
We look forward to a public facing exploration of those models in the future.

Finally, if teams follow our fourth-down recommendations, their behavior will change and win probability will change accordingly.
Statistical win probability models that learn from the game outcomes defining the recent history of football do not account for this distribution shift.
State-space models, on the other hand, can account for these changes by altering the probability that a team goes for in on fourth down as a function of game-state.
A more elaborate analysis would account for this distribution shift.



\if0\blind
{
  \section*{Acknowledgments}

The authors thank the many football analysts who contributed to the development of the fourth down problem. In particular, we thank Brian Burke for providing helpful feedback. The authors acknowledge the High Performance Computing Center (HPCC) at The Wharton School, University of Pennsylvania for providing computational resources that have contributed to the research results reported within this paper. 
} \fi

\bibliography{refs}

\newpage
\begin{center}
{\large\bf SUPPLEMENTARY MATERIAL}
\end{center}
\appendix


\section{Data details}\label{app:data_details}

\begin{table}[H]
\bgroup
\def\arraystretch{1.2}%
\centering
\begin{tabular}{ llllll }
\hline 
variable & variable description \\ \hline
win/loss & 1 if the possession team wins the game, else 0 \\
game seconds remaining & num. seconds remaining in the game in $\{3600,...,1\}$ \\
score differential & \vtop{\hbox{\strut point differential between the offensive and defensive}\hbox{\strut team at the start of this play}} \\
total score & total points scored during this game prior to this play  \\
posteam spread & pre-game Vegas point spread relative to the possession team \\
total points line & pre-game Vegas total points over/under line \\
yards to opp. endzone (i.e., yardline) & \vtop{\hbox{\strut num. yards to the opponent's endzone in}\hbox{\strut $\{0,1,...,99,100\}$ ($0$ is touchdown and $100$ is safety)}}  \\
yards to go (i.e., ydstogo) & \vtop{\hbox{\strut an integer, the number of yards an offense has to gain}\hbox{\strut to achieve a first down or touchdown}} \\
down & a number in $\{1,2,3,4\}$ denoting the down of the play \\
posteam timeouts remaining & num. timeouts in $\{3,2,1,0\}$ the offensive team has  \\
defteam timeouts remaining & num. timeouts in $\{3,2,1,0\}$ the defensive team has \\
receive 2h ko & 1 if the possession team receives the $2^{nd}$ half kickoff, else 0 \\
home & 1 if the offensive team is at home, else 0 \\
era & \vtop{\hbox{\strut categorical variable grouping the year into}\hbox{\strut$\{1999-2005, \ 2006-2013, \ 2014-2017, \ 2018-2022\}$  }} \\
roof & \vtop{\hbox{\strut categorical variable grouping the roof of the stadium into}\hbox{\strut$\{ \text{closed, dome, open, outdoors} \}$  }} \\
game id & unique identifier of this game \\
drive id & unique identifier of this drive \\
posteam coach & name of the coach of the possession team \\
\hline 
\end{tabular}
\caption{
Game-state variables relevant to estimating win probability that describe the context at the start of a play.
}
\label{table:wp_variables}
\egroup
\end{table}

\section{Estimating player / team quality}\label{app:player_quality}

\textbf{Kicker quality.} 
We define a kicker's quality by a weighted mean of his field goal probability added over all his previous kicks in his career.
To begin, we fit a simple kicker-agnostic field goal probability model $\fgpzero$ using logistic regression as a function of yardline (specifically, a cubic polynomial in yardline). 
Then, we define the \textit{field goal probability added} ($\fgpa$) of the $n^{th}$ field goal by
\begin{equation}
    \fgpa_n := \ind{n^{th} \text{ field goal is made}} - \fgpzero(\text{yardline of the } n^{th} \text{ field goal}).
\end{equation}
Now we define kicker quality using the Ravens' kicker Justin Tucker for concreteness.
Index all of Tucker's field goals by $n$. 
We define Tucker's kicker quality prior to field goal $n \geq 2$ by a weighted mean of the field goal probability added in his previous kicks,
\begin{equation}
    \kq_n := \frac{\sum_{j=1}^{n-1} \alpha^{n-1-j} \cdot \fgpa_j}{\gamma + \sum_{j=1}^{n-1} \alpha^{n-1-j}},
\end{equation}
and $\kq_1 := 0$.
The hyperparameter $\gamma$ stabilizes $\kq$ for low $n$, shrinking $\kq$ towards 0 (i.e., shrinking towards an average kicker).
In other words, we impute $\gamma$ synthetic field goals, each with an ``average'' $\fgpa$ outcome of zero.
We use $\gamma = 96$.
The hyperparameter $\alpha$ is an exponential decay weight, which upweights more recent kicks, allowing us to capture non-stationarity in kicker quality over time.
We use $\alpha = 0.985$. 
For instance, a kick which occurred 46 kicks ago is weighted half as much as the previous kick since $0.985^{46-1} \approx 0.5$. 
Also, a kick which occurred 153 kicks ago is weighted one-tenth as much as the previous kick since $0.985^{153-1} \approx 0.1$.
Finally, we standardize kicker quality to have mean zero and standard deviation $1$.

\begin{figure}[hbt!]
    \centering{}
    {\includegraphics[width=0.8\textwidth]{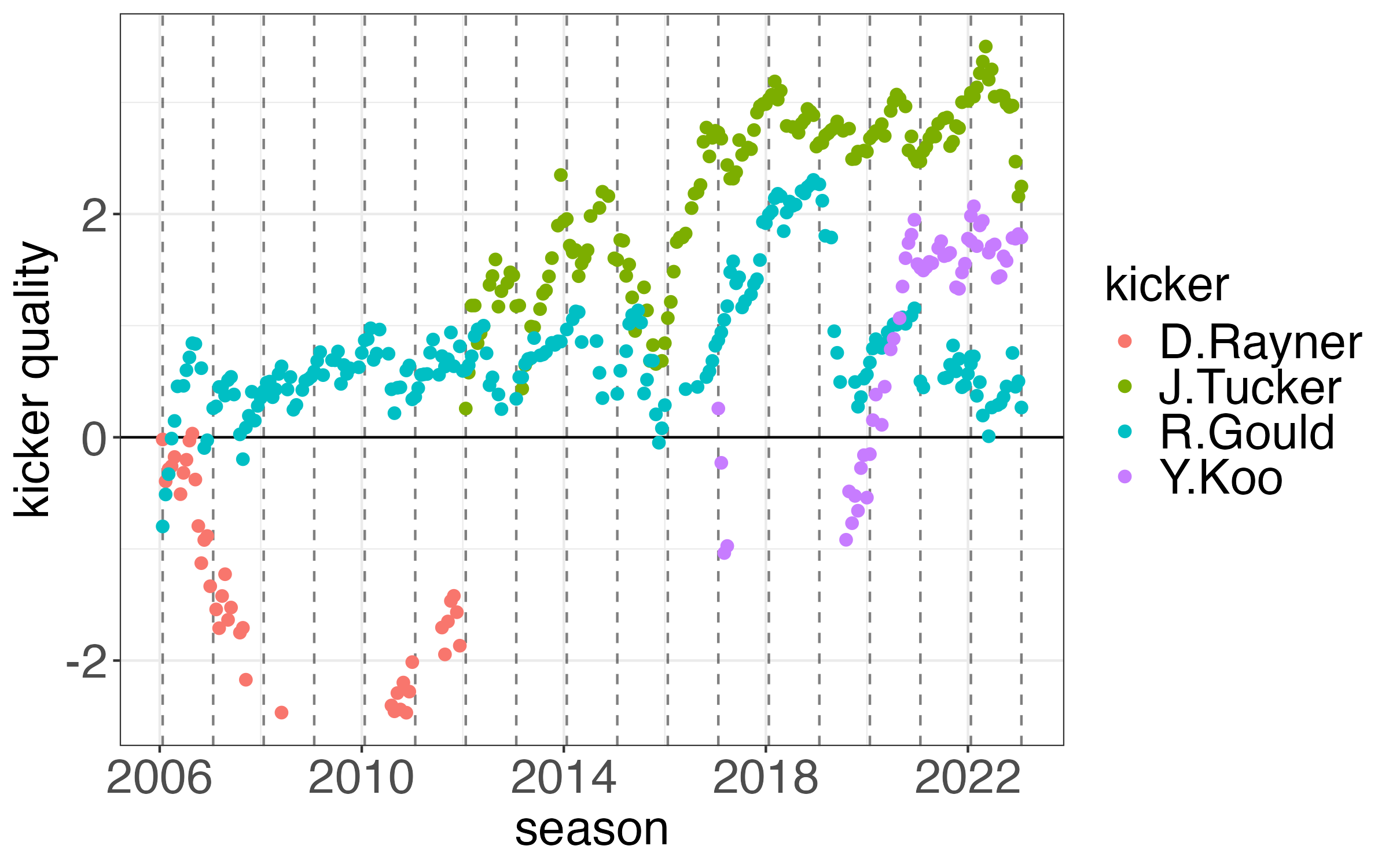} }
    \caption{
    Kicker quality trajectories versus time for various kickers.
    }
    \label{fig:kq_trajs}
\end{figure}

\begin{figure}[hbt!]
    \centering{}
    \subfloat[\centering \label{fig:kq_rankings}]{{\includegraphics[width=0.5\textwidth]{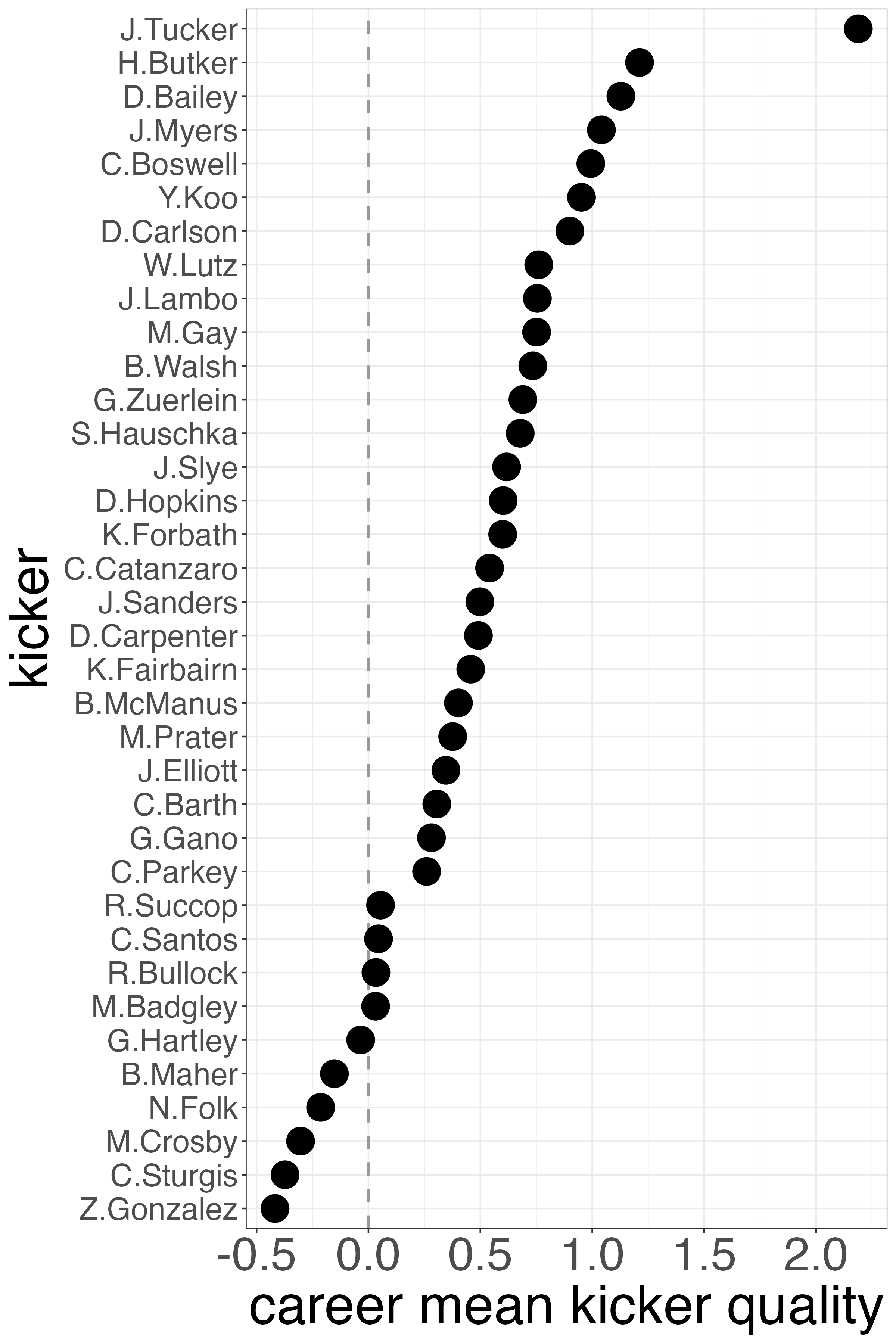} }}%
    \subfloat[\centering \label{fig:pq_rankings}]{{\includegraphics[width=0.5\textwidth]{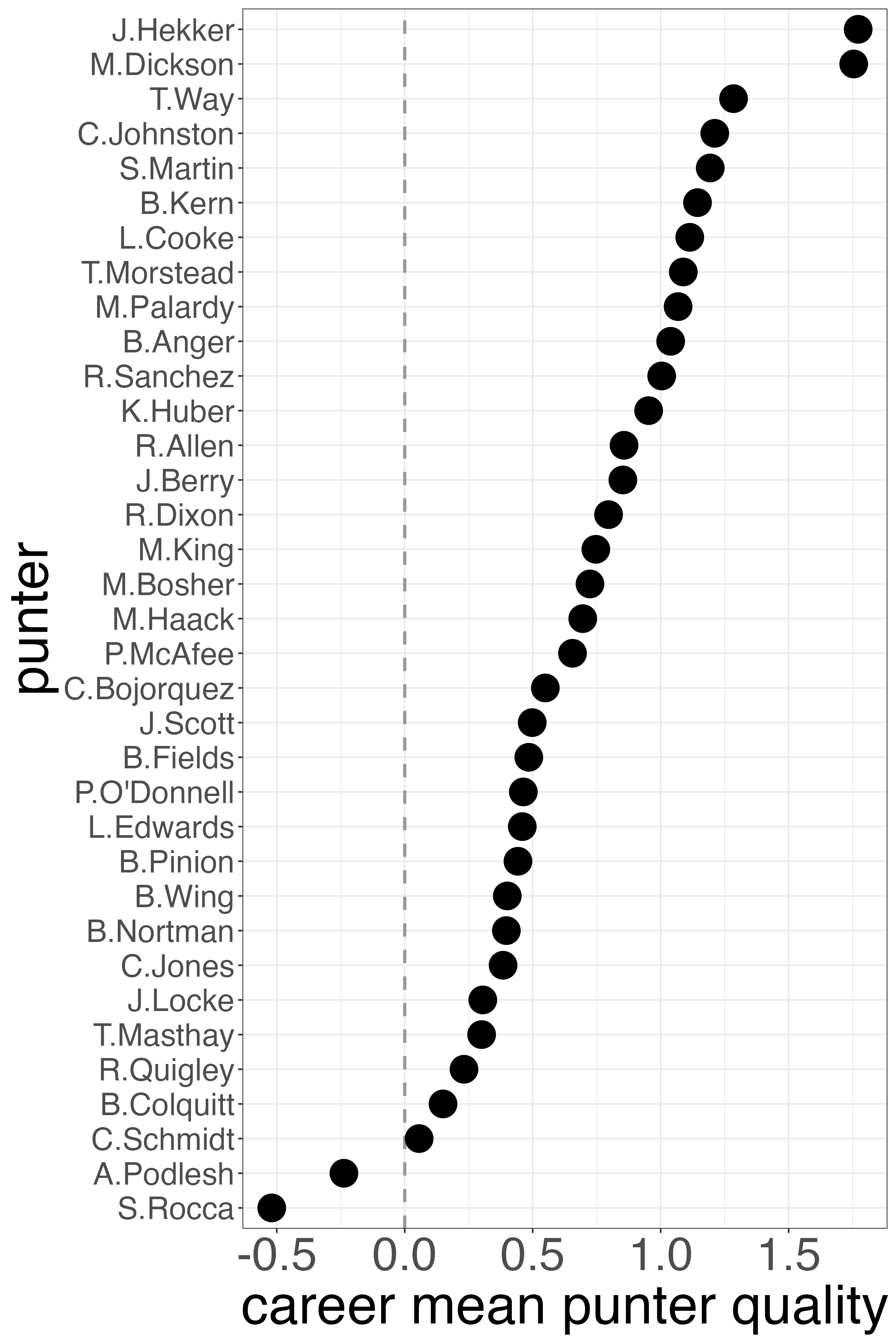} }}%
    \caption{
    (a) The career mean kicker quality ($x$-axis) of each kicker ($y$-axis) with over 100 field goal attempts  in our dataset.
    (b) The career mean punter quality ($x$-axis) of each punter ($y$-axis) with over 250 punts  in our dataset. 
    }
\end{figure}

In Figure~\ref{fig:kq_trajs} we visualize the kicker quality trajectories of a few kickers across their careers.
We see non-stationarity in kicker quality over time.
For example, notice Younghoe Koo's sharp rise in 2020 (he made the Pro Bowl that year) and Robbie Gould's dip in 2019.
In Figure~\ref{fig:kq_rankings} we plot the career mean kicker quality of each kicker with over 100 field goal attempts in our dataset.
As expected, Justin Tucker has by far the highest kicker quality.

\textbf{Punter quality.} 
We define punter quality in a similar fashion as kicker quality.
In particular, we define a punter's quality by a weighted mean of his punt yards over expected over all his previous punts in his career.
To begin, we fit a simple punter-agnostic expected next yardline after punting model $\enyzero$ using linear regression as a function of yardline (specifically, a cubic polynomial in yardline). 
Then, we define the \textit{punt yards over expected} ($\pyoe$) of the $n^{th}$ punt by
\begin{equation}
    \pyoe_n := \text{actual yardline after the } n^{th} \text{ punt} - \enyzero(\text{yardline prior to the } n^{th} \text{ punt}).
\end{equation}
Now we define punter quality using Rams' punter Johnny Hekker for concreteness.
Index all of Hekker's punts by $n$. 
We define Hekker's punter quality prior to punt $n \geq 2$ by a weighted mean of the punt yards over expected in his previous kicks,
\begin{equation}
    \pq_n := \frac{\sum_{j=1}^{n-1} \alpha^{n-1-j} \cdot \pyoe_j}{\gamma + \sum_{j=1}^{n-1} \alpha^{n-1-j}},
\end{equation}
and $\pq_1 := 0$.
The hyperparameters $\gamma$ and $\alpha$ play the same role as in kicker quality.
We use $\gamma = 150$ and $\alpha = 0.99$.
Finally, we standardize punter quality to have mean zero and standard deviation $1$.

\begin{figure}[hbt!]
    \centering{}
    {\includegraphics[width=0.8\textwidth]{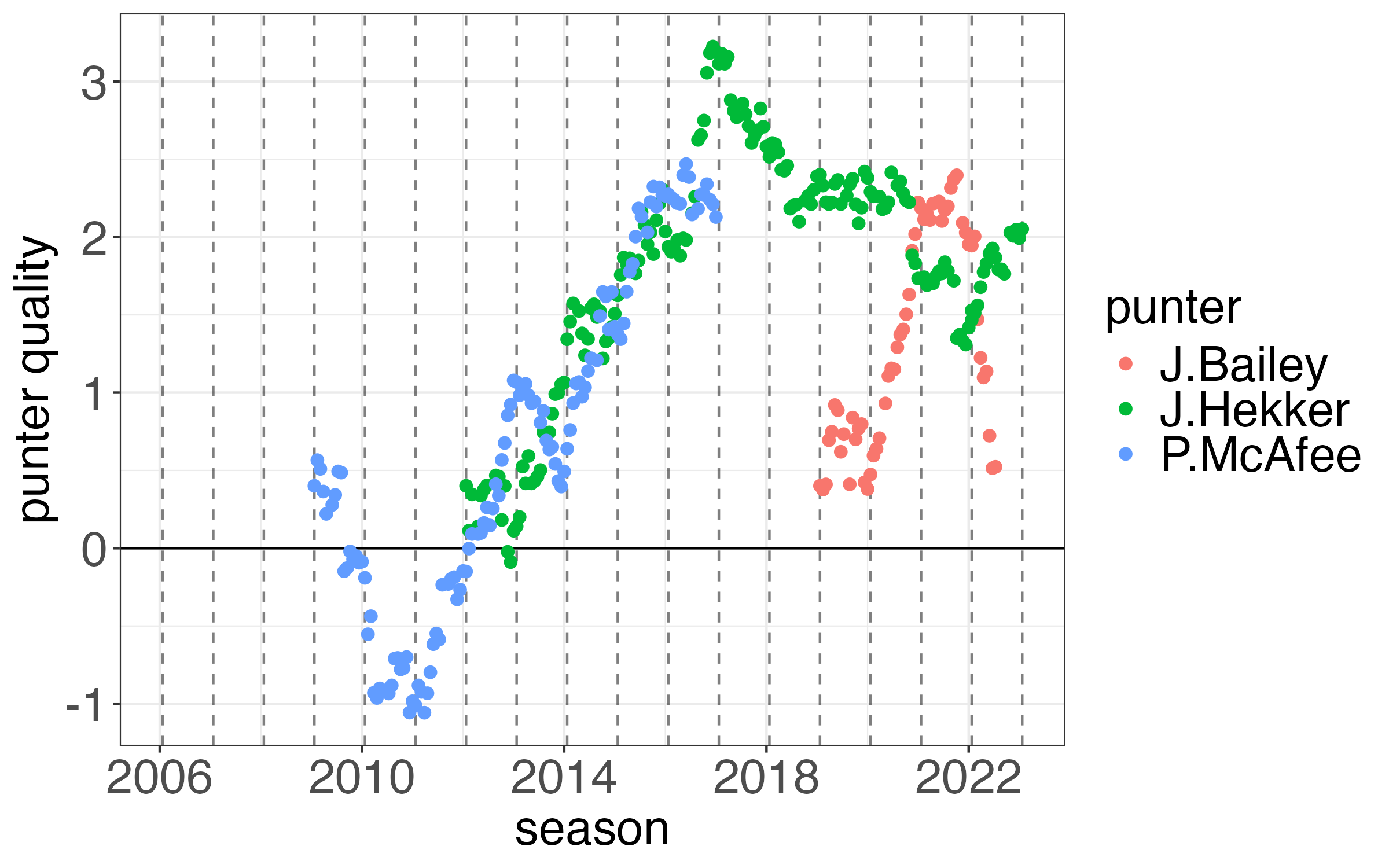} }
    \caption{
    Punter quality trajectories versus time for various punters.
    }
    \label{fig:pq_trajs}
\end{figure}

In Figure~\ref{fig:pq_trajs} we visualize the punter quality trajectories of a few punters across their careers.
We see that non-stationarity in punter quality over time.
For example, notice Pat McAfee's steady rise across his mid and later years (he made the Pro Bowl in 2014 and 2016) and Jake Bailey's decline in 2022.
In Figure~\ref{fig:pq_rankings} we plot the career mean punter quality of each punter with over 250 punts in our dataset.
As expected, Johnny Hekker has the highest punter quality.

\textbf{Difference in offensive and defensive quality.} 
We create a measure of how much better the offensive team's offensive quality is than the defensive team's defensive quality using information from betting markets.
We use Vegas' pre-game point spread relative to the offensive team (denoted $\ps$) and the total points over/under line (denoted $\tp$).
In particular, our measure is $(\tp - \ps)/2$, standardized to have mean zero and standard deviation $1$.

We derive this expression as follows. 
Denote the offensive quality of the offensive team by $\oqot$, the defensive quality of the defensive team by $\dqdt$, the offensive quality of the defensive team by $\oqdt$, and the defensive quality of the offensive team by $\dqot$.
These variables are not observable, and we are interested in the differences $\oqot - \dqdt$ and $\oqdt - \dqot$.
The point spread is a measure of how much better the offensive team's offense's edge is over the opposing defense relative to the opposing team's edge.
A reasonable model is
\begin{equation}
\ps = -((\oqot - \dqdt) - (\oqdt - \dqot)),
\end{equation}
recalling that, by convention, negative point spreads indicate larger edges.
The total points line is a measure of the combined strenth of both offense's edges.
A reasonable model is
\begin{equation}
\tp = (\oqot - \dqdt) + (\oqdt - \dqot) + \overline{\tp},
\end{equation}
where $\overline{\tp}$ denotes the average total points line if neither offense has an edge.
From these two equations, we derive $\oqot - \dqdt = (\tp - \ps)/2$ and $\oqdt - \dqot = (\tp + \ps)/2$, ignoring an additive constant.

\section{Tuning the number of bootstrapped datasets}\label{app:choosing_B}

We use $\bootp$ to measure uncertainty in the estimated optimal decision, but we cannot necessarily rely on a non-technical football coach to fully process how $\bootp$ maps to the strength of a recommendation.
To facilitate easier communication to a coach, we bin decision recommendations into three buckets: confident ($\bootp \in [83\%, 100\%]$), lean ($\bootp \in [67\%, 83\%)$), and uncertain ($\bootp \in [0\%, 67\%)$).
Since we ultimately bucket decision recommendations into these three bins (see Section~\ref{sec:overconfidence}), we want the number of bootstrapped models $B$ to be large enough to stably categorize decisions into one of these three bins.
We also want $B$ to be as small as possible in order to quickly evaluate a fourth-down recommendation during a football game.
Hence, we conduct a stability analysis to choose $B$.

\begin{figure}[hbt!]
    \centering
    \includegraphics[width=0.7\textwidth]{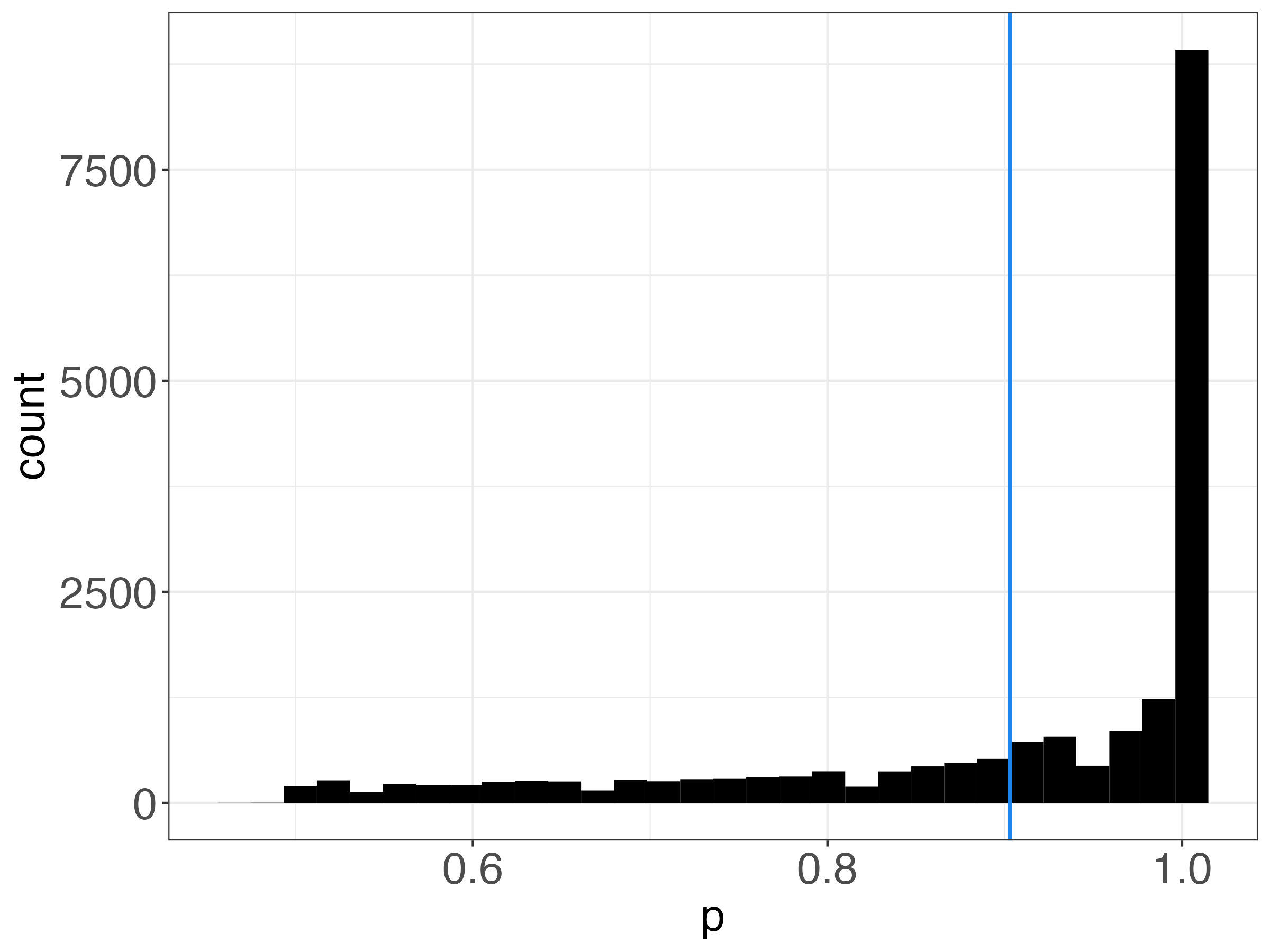}
    \caption{
        This is a histogram of $\{p_{i}^{(B)}\}_i$. The blue line is the mean. Since most of the values equal 1 and the mean is high enough, $B=101$ is large enough.
    }
    \label{fig:stability_analysis}
\end{figure}

Across $M=100$ draws of $B$ bootstrapped decision models in each, we categorize each observed fourth-down play from 2018 to 2022 into one of the three bins.
Let $i$ index all the observed fourth-down plays from 2018 to 2022, let $m \in \{1,...,M\}$ index the draw of the bootstrap, and in each draw of the bootstrap we generate $B$ fourth-down decision models.
Given $\bootp_{im}^{(B)}$, the $\bootp$ of the estimated optimal decision for play $i$ in draw $m$ of the bootstrap, we calculate
\begin{equation}
\label{eqn:boot_stability}
\begin{cases}
    p_{i}^{\text{confident, }(B)} = \frac{1}{M} \sum_{m=1}^{M} \ind{\bootp_{im}^{(B)} \in [83\%, 100\%] }, \\
    p_{i}^{\text{lean, }(B)} = \frac{1}{M} \sum_{m=1}^{M} \ind{\bootp_{im}^{(B)} \in [67\%, 83\%) }, \\
    p_{i}^{\text{uncertain, }(B)} = \frac{1}{M} \sum_{m=1}^{M} \ind{\bootp_{im}^{(B)} \in [0\%, 67\%) }, \\
    p_{i}^{(B)} = \max\{p_{i}^{\text{confident, }(B)}, \ p_{i}^{\text{lean, }(B)}, \ p_{i}^{\text{uncertain, }(B)}\}.
\end{cases}
\end{equation}
For each category, $p_{i}^{\text{category, }(B)}$ is the proportion of the $M$ bootstrap draws that play $i$ is put in that category.
For our procedure with $B$ bootstrapped models to be stable, we want $p_{i}^{\text{category, }(B)}$ to be close to 1 for each category, which means that a play's categorization is not dependent on the randomness inherent in generating $B$ bootstrapped models.
Hence, we want $p_{i}^{(B)}$, the maximum $p_{i}^{\text{category, }(B)}$ across the three categories, to be close to 1.
Hence, we want the mean across all plays $\overline{p}^{(B)} = \frac{1}{n}\sum_{i=1}^{n} p_{i}^{(B)}$ to be close to 1.
In Figure~\ref{fig:stability_analysis} we see that for $B=101$, $\overline{p}^{(B)} = 0.9$ and the vast majority of plays $i$ have $p_{i}^{(B)} = 1$.
Some plays have $p_{i}^{(B)}$ lower than 1 since they lie near the border of two categories.
We believe $\overline{p}^{(B)} = 0.9$ is sufficiently large and hence use $B=101$ in our fourth-down decision procedure.

\section{Baseline coaches' decision model}\label{app:baseline_coach_model}

\begin{figure}[hbt!]
    \centering{}
    \subfloat[\centering \label{fig:coach_ex_1}]{{\includegraphics[width=7.5cm]{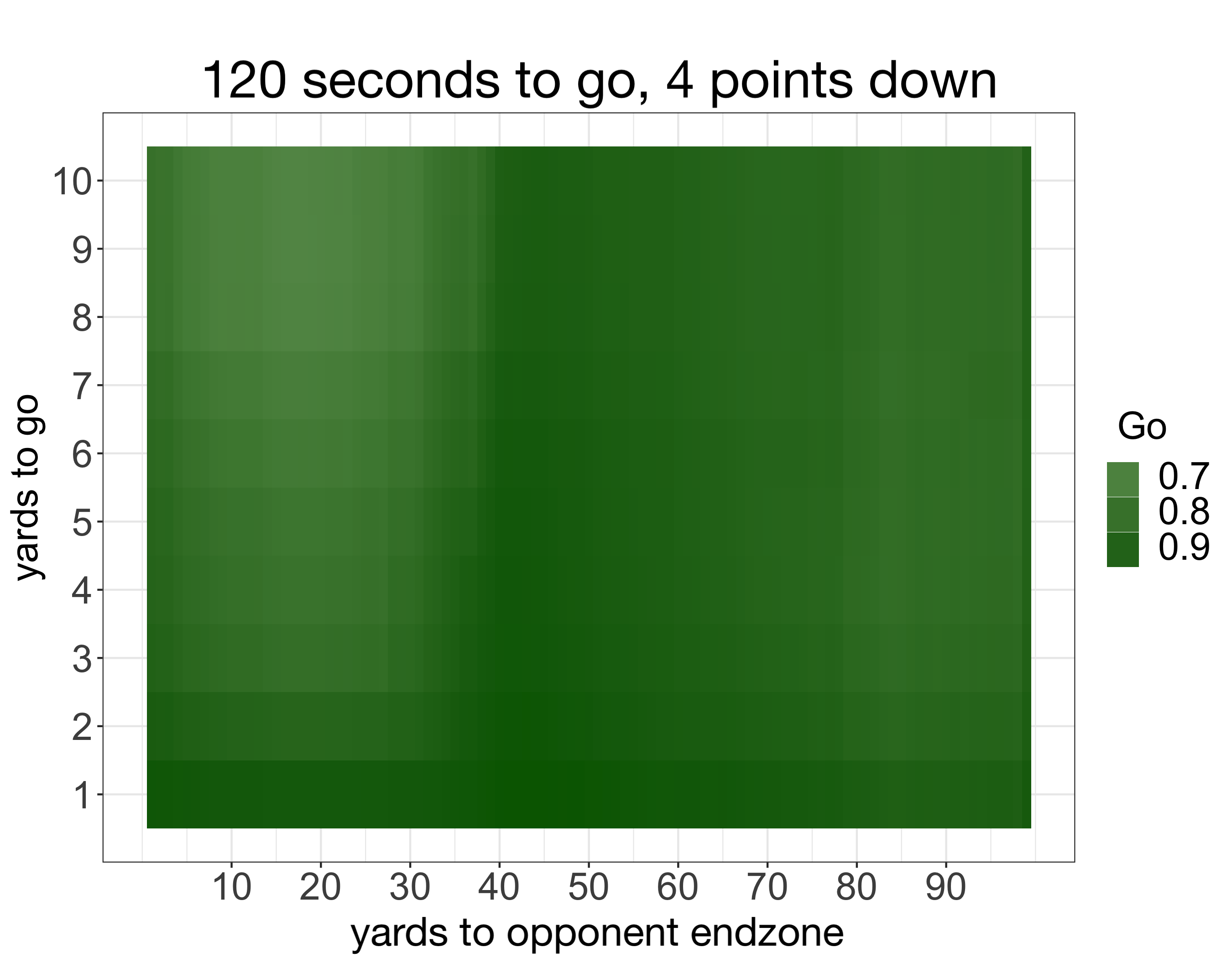} }}%
    \qquad
    \subfloat[\centering \label{fig:coach_ex_2}]{{\includegraphics[width=7.5cm]{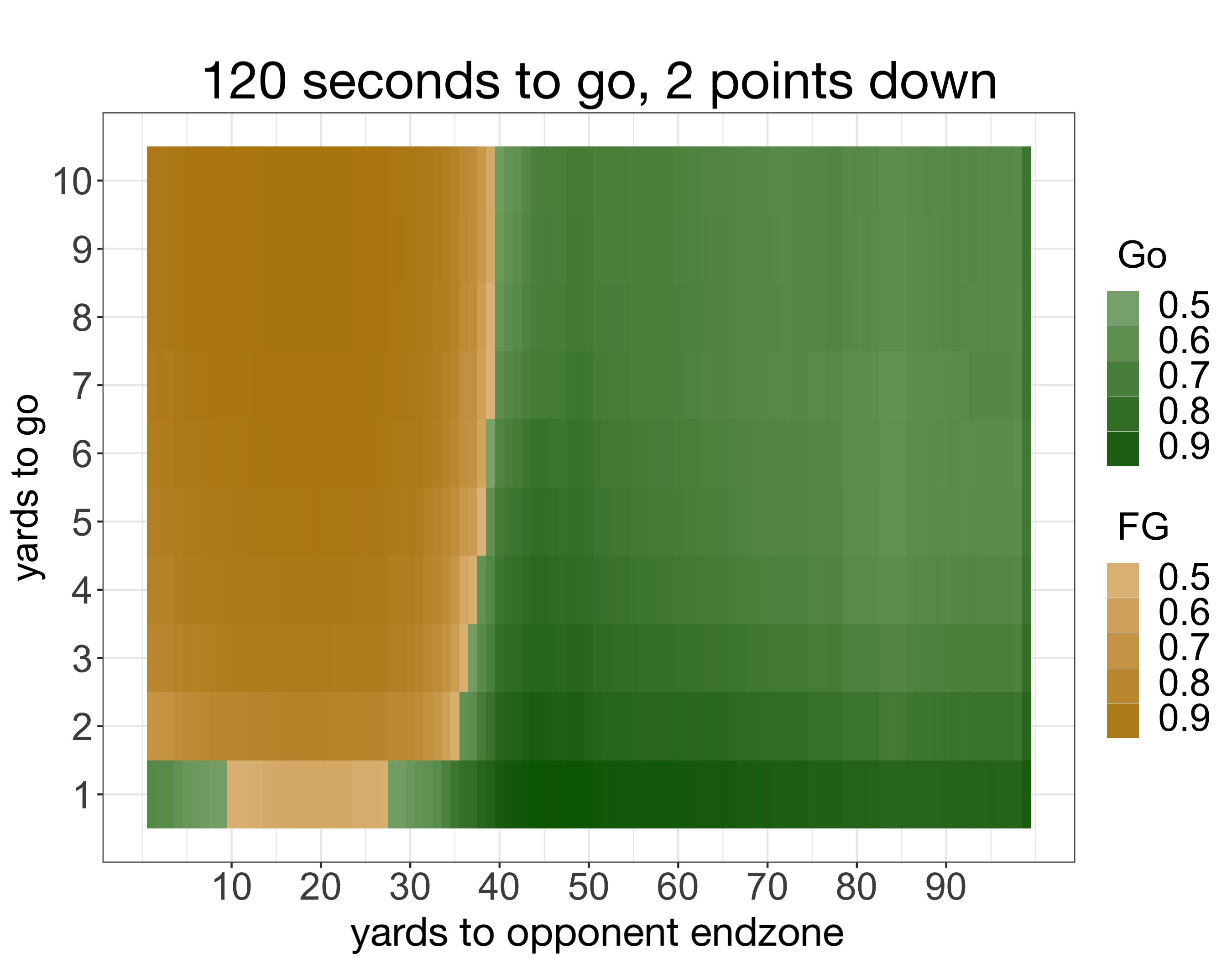} }}
    \qquad
    \subfloat[\centering \label{fig:coach_ex_3}]{{\includegraphics[width=7.5cm]{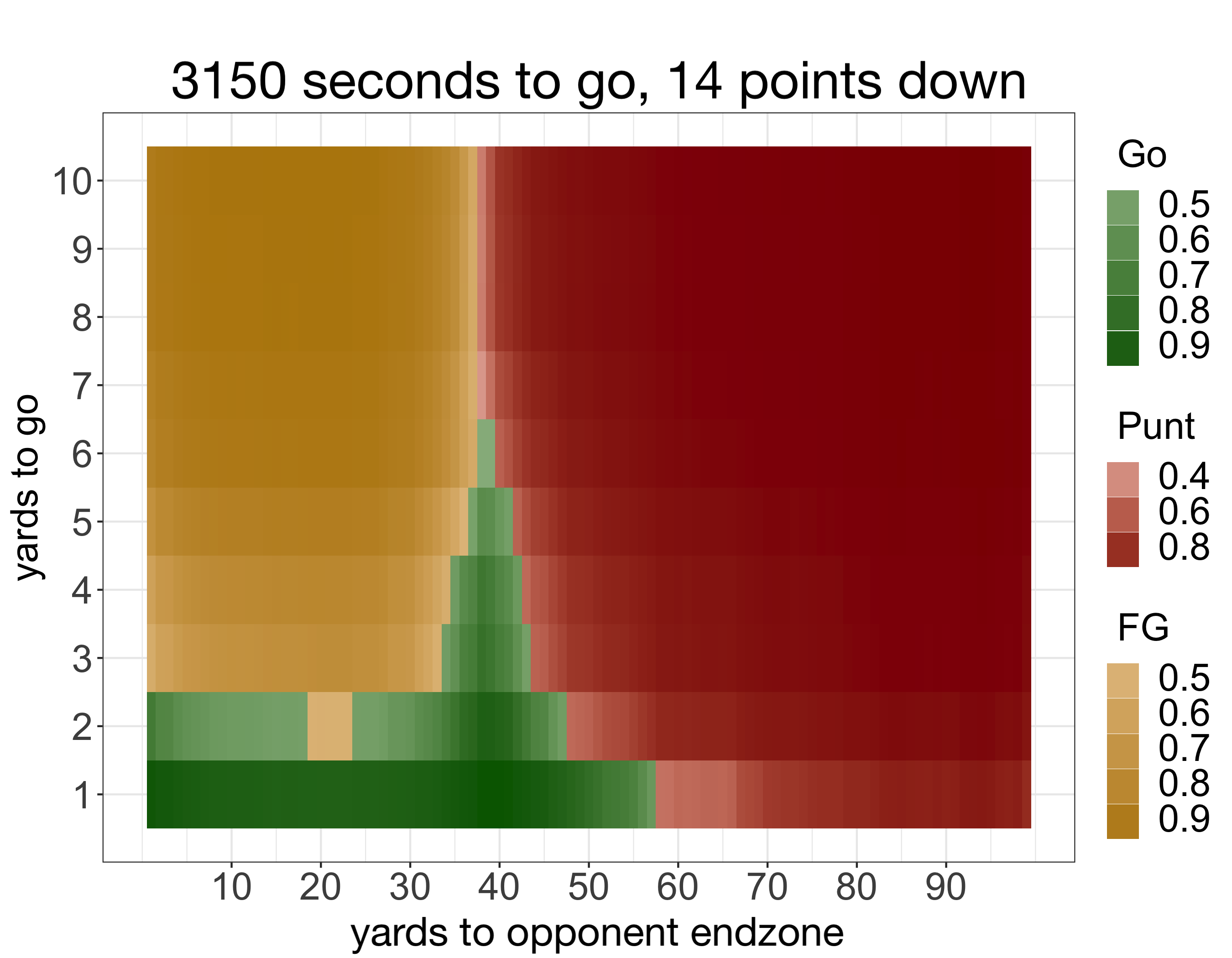} }}
    \qquad
    \subfloat[\centering \label{fig:coach_ex_4}]{{\includegraphics[width=7.5cm]{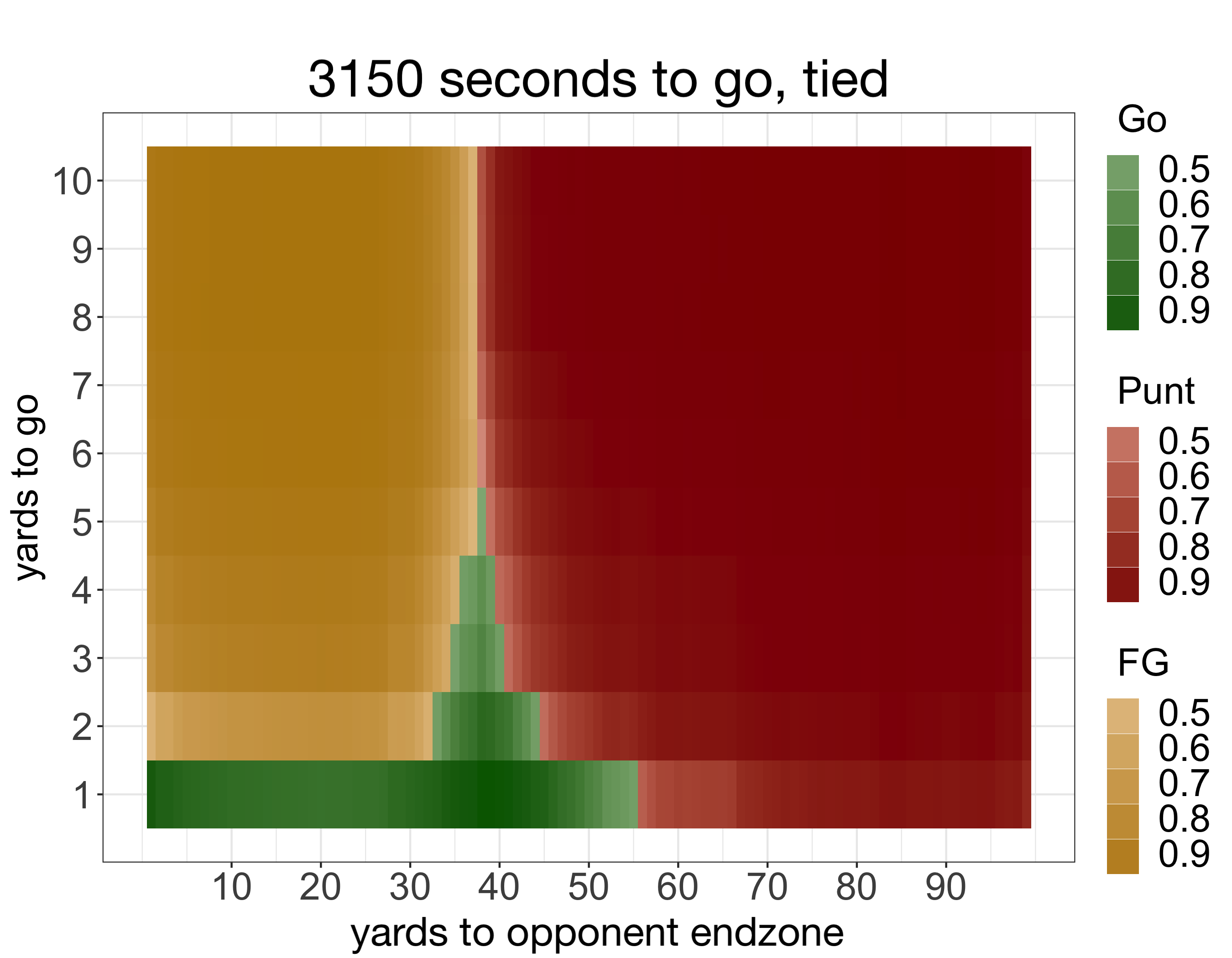} }}
    \caption{
    Visualizing our model of the typical coach's fourth-down decision as a function of yards to opponent endzone and yards to go, for various values of time remaining and score differential. Green, yellow, and red indicate that $\go$, $\fg$, and $\punt$ is the most likely decision, respectively.
    The color intensity reflects the likelihood that a coach makes that decision.
    }
    \label{fig:coach_model_viz}
\end{figure}

To compare our decision-making procedure to the decisions that actual football coaches tend to make, we model the probability that a coach chooses a decision in $\{\go,\fg,\punt\}$ as a function of game-state.
We use $\xgb$ to fit these coach probabilities.
$\xgb$ works well here because we have $94,786$ fourth-down plays in our full dataset of plays since 1999, and each play is a reasonably approximate independent observation of a coach's decision.
In particular, we fit these coach probabilities as a function of yards to opponent endzone, yards to go, game seconds remaining, score differential, point spread, and era (1999-2001, 2002-2005, 2006-2013, 2014-2017, and 2018-present).
In Figure~\ref{fig:coach_model_viz} we visualize these coach decision models, and the results make intuitive sense.
For the most part, coaches punt deep in their own territory and kick field goals near the opponent's endzone, except for with one and sometimes two yards to go.
Also, at the end of the game, coaches' decision-making changes depending on the number of points they need to score to win the game.

\begin{figure}[hbt!]
    \centering{}
    {\includegraphics[width=11cm]{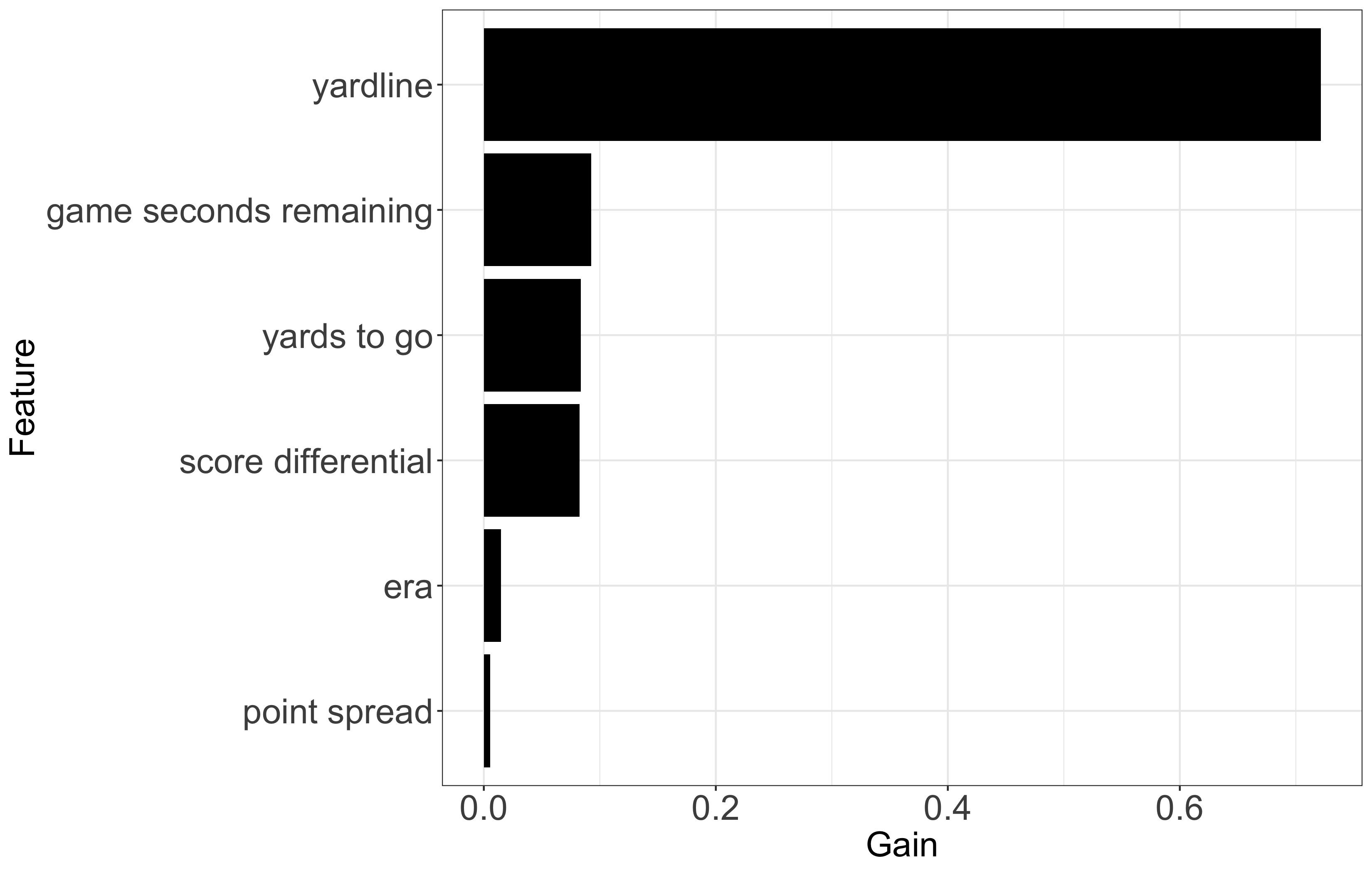} }
    \caption{Variable importance (gain) for the typical coach's decision probability model.}
    \label{fig:plot_coach_xgb_importance}
\end{figure}

In Figure~\ref{fig:plot_coach_xgb_importance} we visualize the variable importance (via gain) of our $\xgb$ model. 
Interestingly, point spread has an extremely small impact on coaches' fourth-down decisions.
We find, however, that point spread should impact fourth-down decision making.
For instance, in certain game-states, it is advantageous for the favorites to be more aggressive (e.g., late in close games).

\section{Additional example plays}\label{app:additional_ex_plays}

\begin{figure}[hbt!]
    \centering
    \includegraphics[width=1\textwidth]{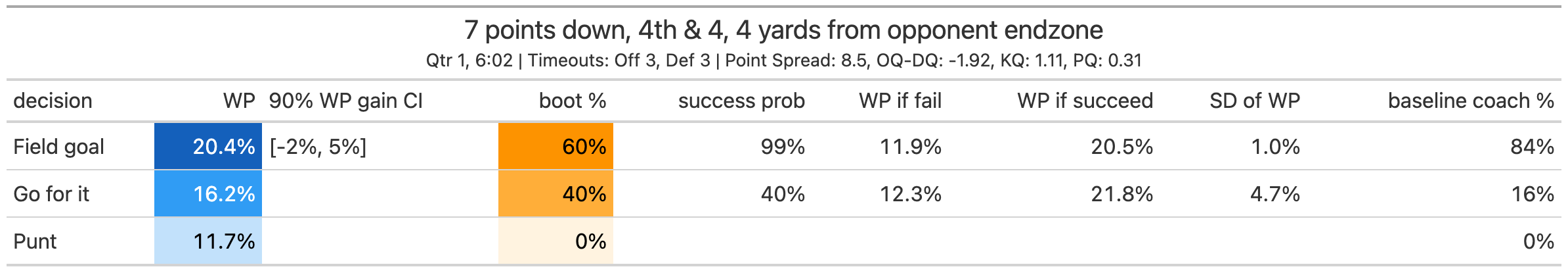}
    \begin{minipage}{\textwidth}
        \centering
        \includegraphics[width=0.48\textwidth]{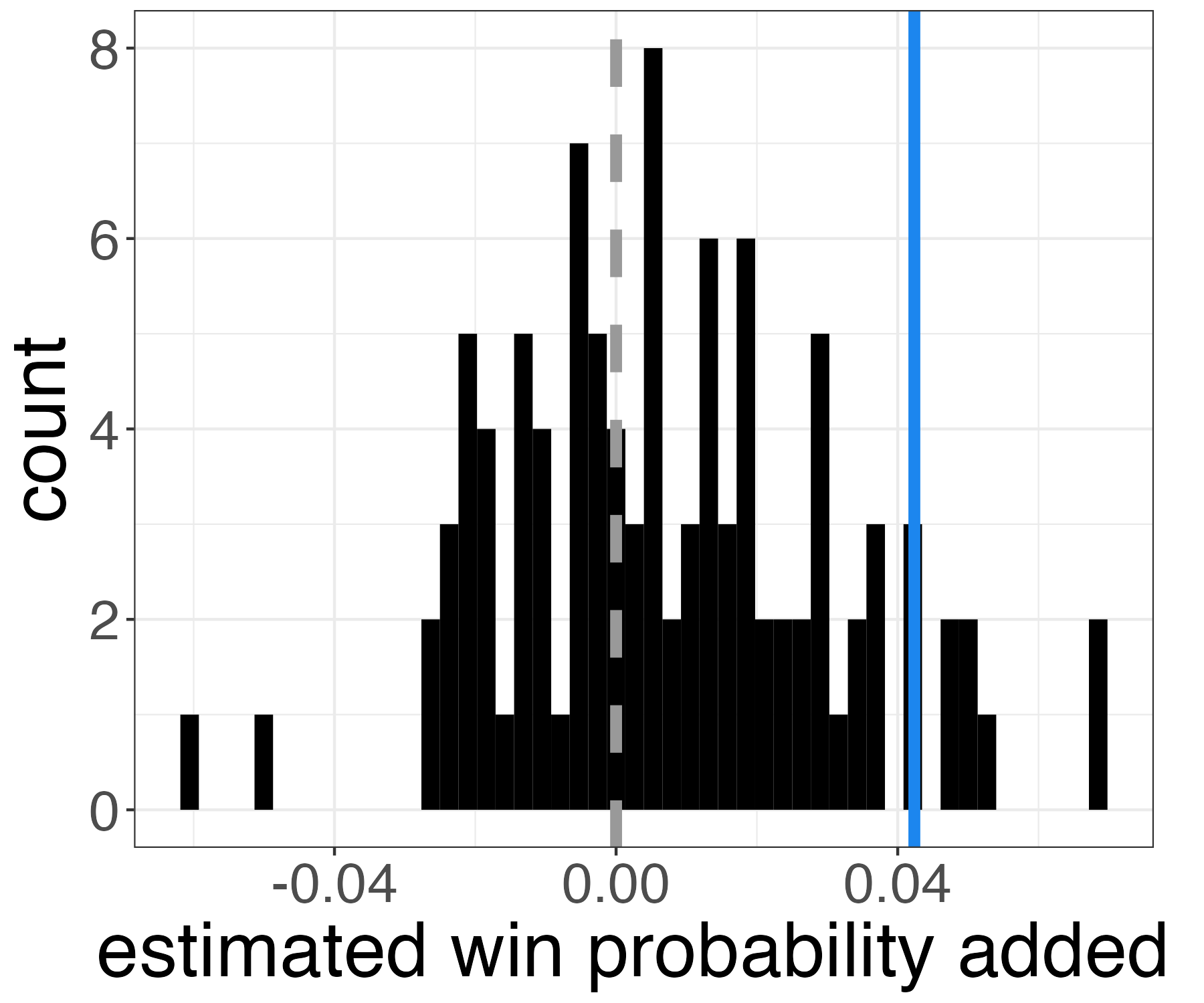}
    \end{minipage}
    \begin{minipage}{\textwidth}
        \centering
        \includegraphics[width=0.48\textwidth]{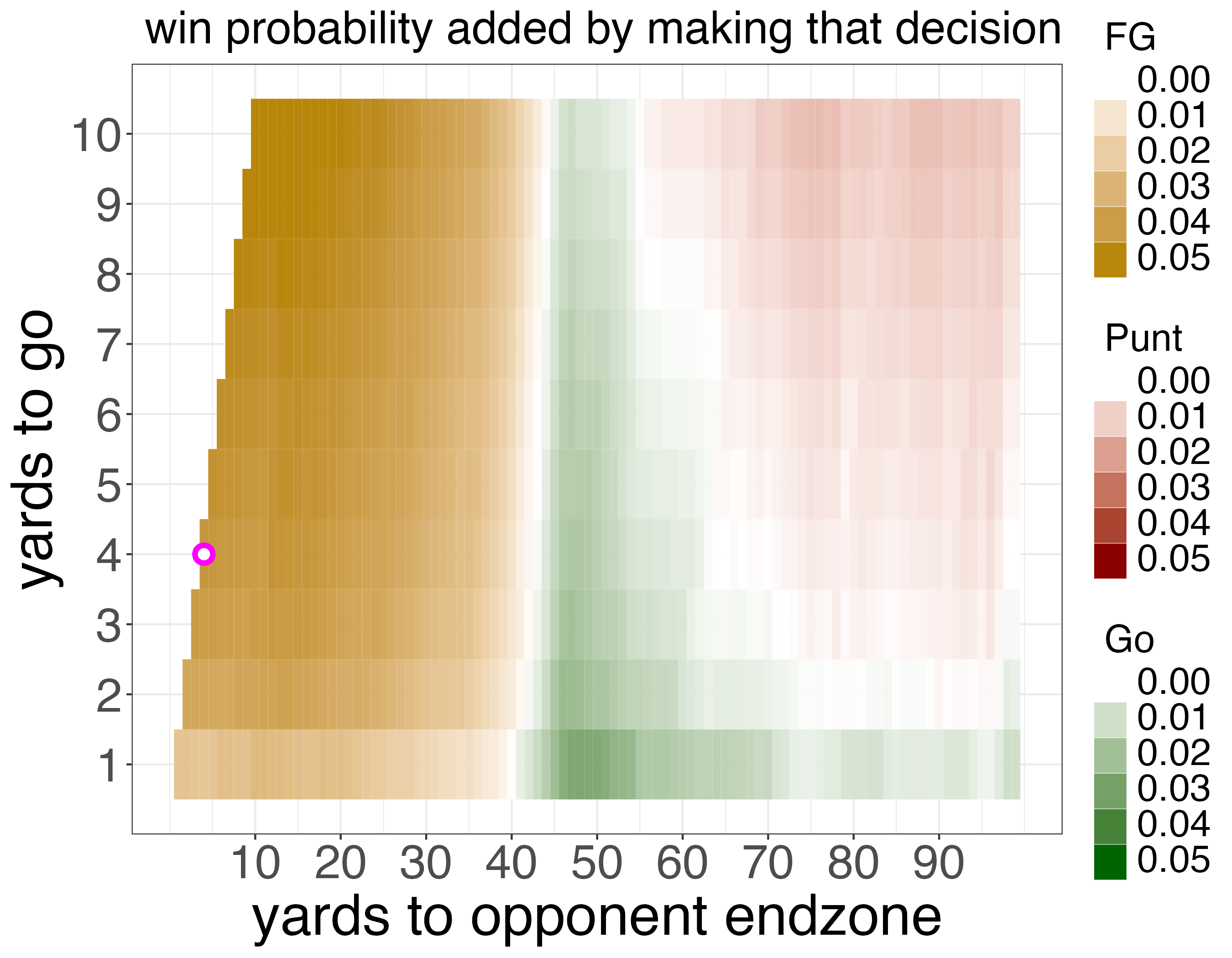}
        \includegraphics[width=0.48\textwidth]{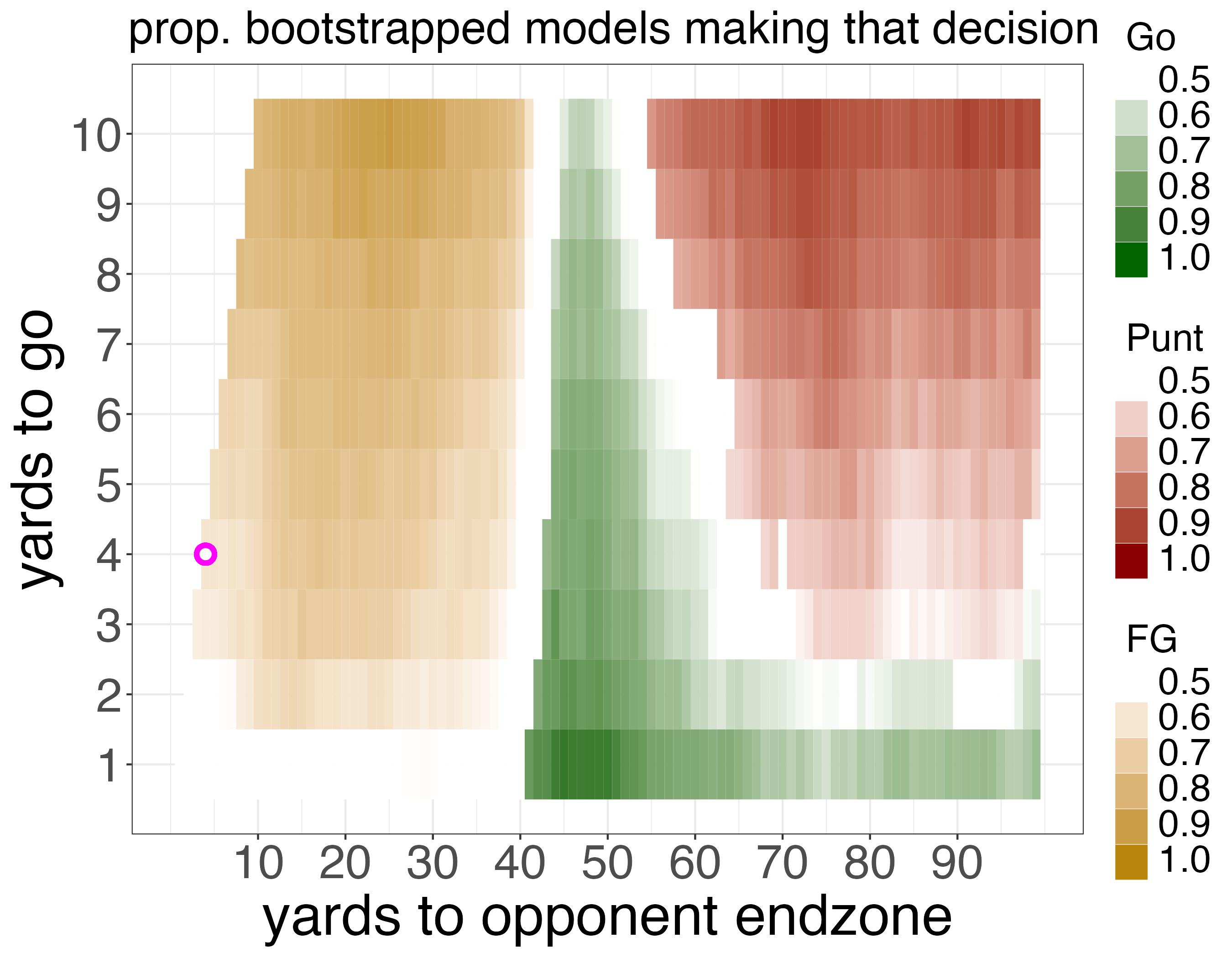}
    \end{minipage}
    \caption{
        Decision charts for example play 3.
    }
    \label{fig:ex_play_3}
\end{figure}

\textbf{Example play 3.} In Figure~\ref{fig:ex_play_3} we visualize the decision procedure for a fourth-down play in which the Bears had the ball against the Jets in Week 12 of 2022.
$\fg$ provides an edge over $\go$ according to the $\wp$ point estimate ($+4.2\%$ $\wp$).
But, the $90\%$ confidence interval of the estimated gain in win probability by attempting a field goal is $[-2\%, 5\%]$, indicating that $\fg$ could either be a good or a bad decision.
Also, $40\%$ of the bootstrapped models say $\go$ is better.
In other words, we do not have enough data to be confident in the win probability point estimates, and we don't know the optimal fourth-down decision at this game-state.
Further, in the bottom right plot, notice how most of the colors are light.
This indicates that the optimal decision is uncertain at most other combinations of yardline and yards to go at this game-state.

\begin{figure}[hbt!]
    \centering
    \includegraphics[width=1\textwidth]{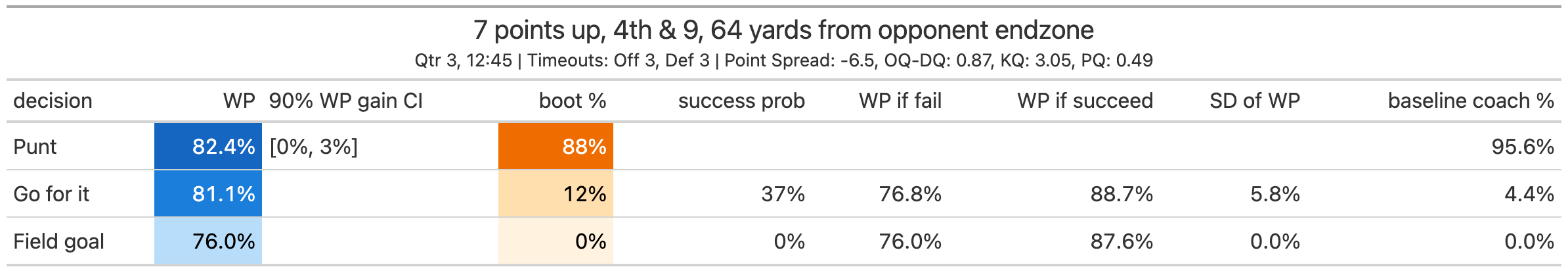}
    \begin{minipage}{\textwidth}
        \centering
        \includegraphics[width=0.48\textwidth]{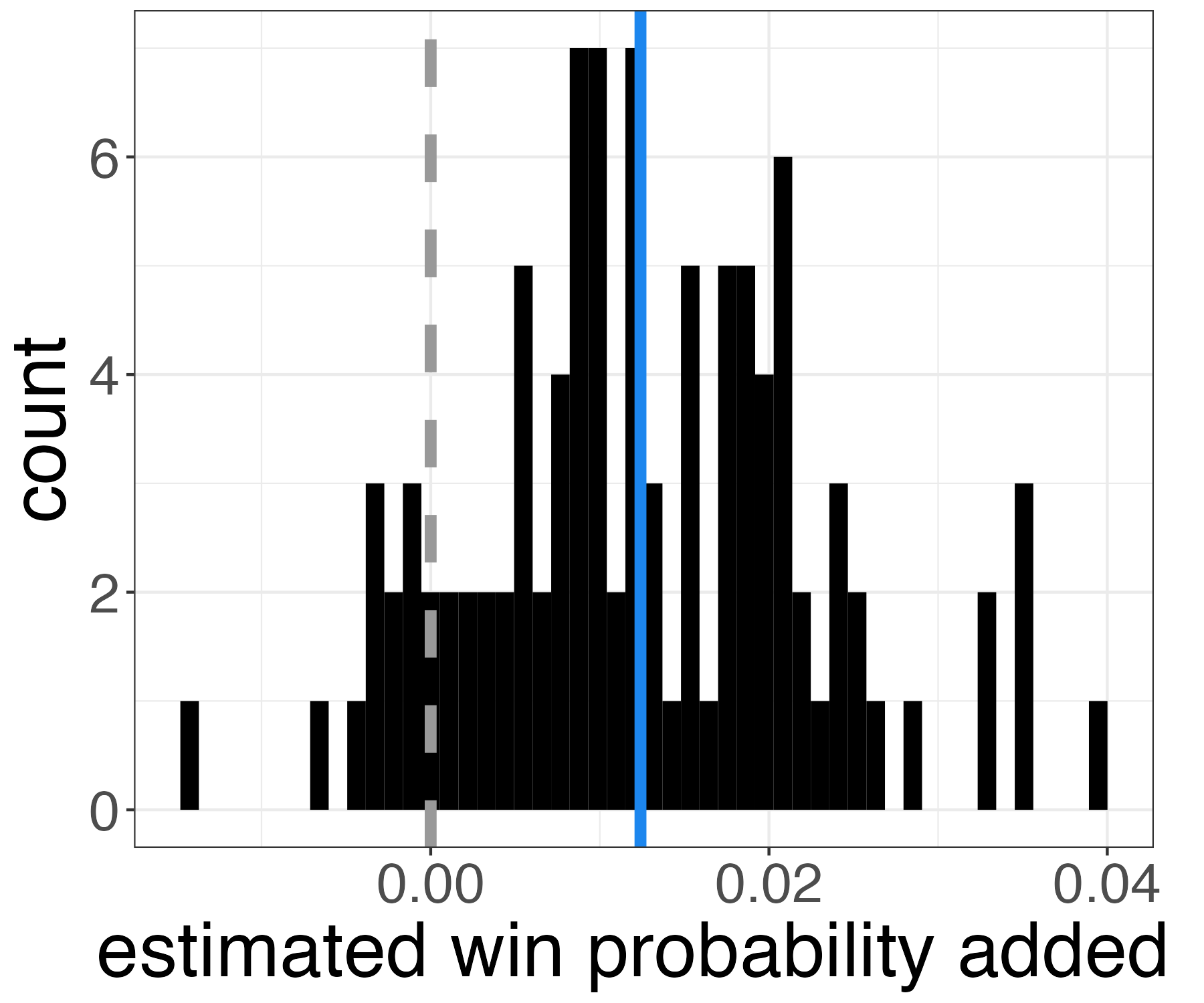}
    \end{minipage}
    \begin{minipage}{\textwidth}
        \centering
        \includegraphics[width=0.48\textwidth]{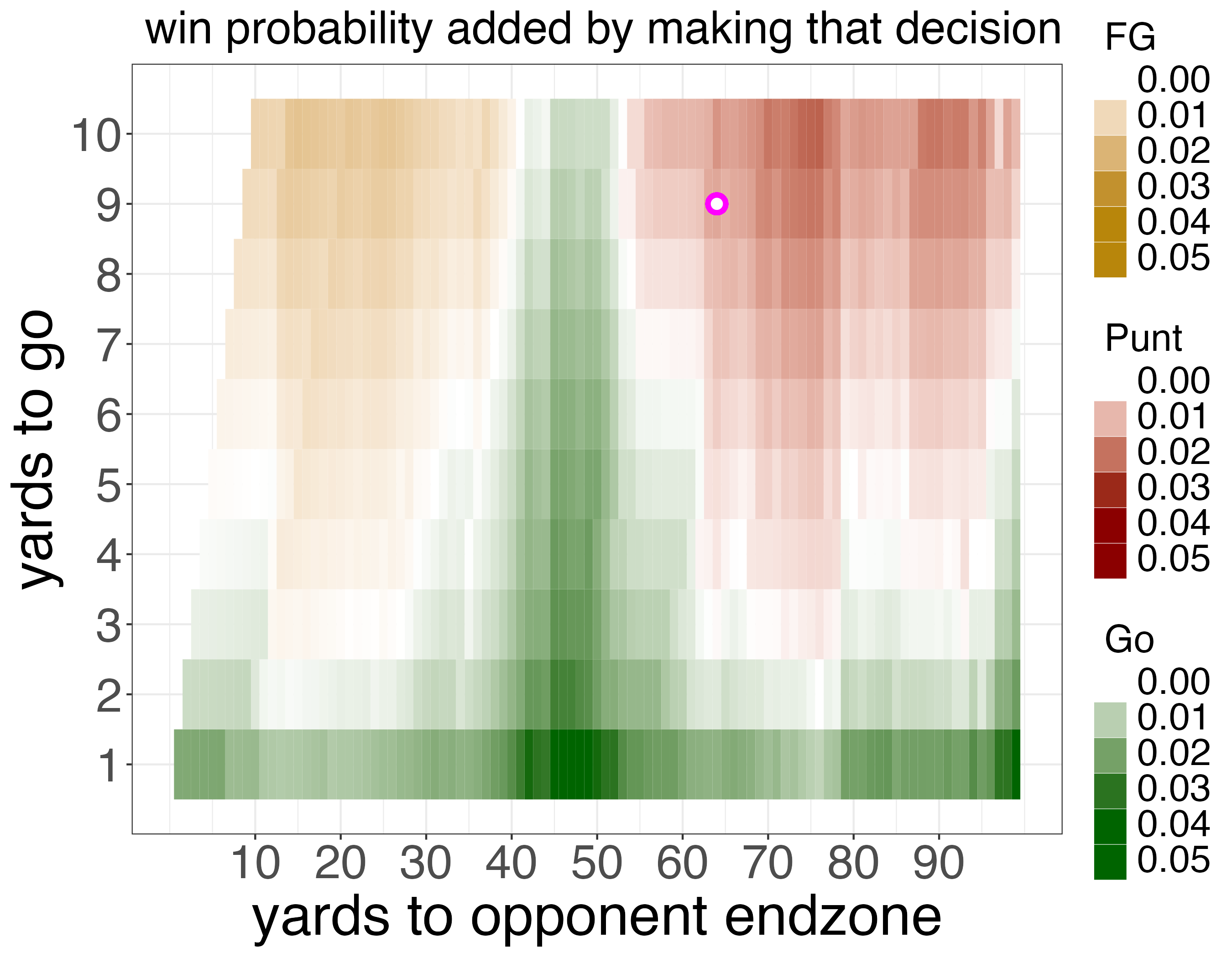}
        \includegraphics[width=0.48\textwidth]{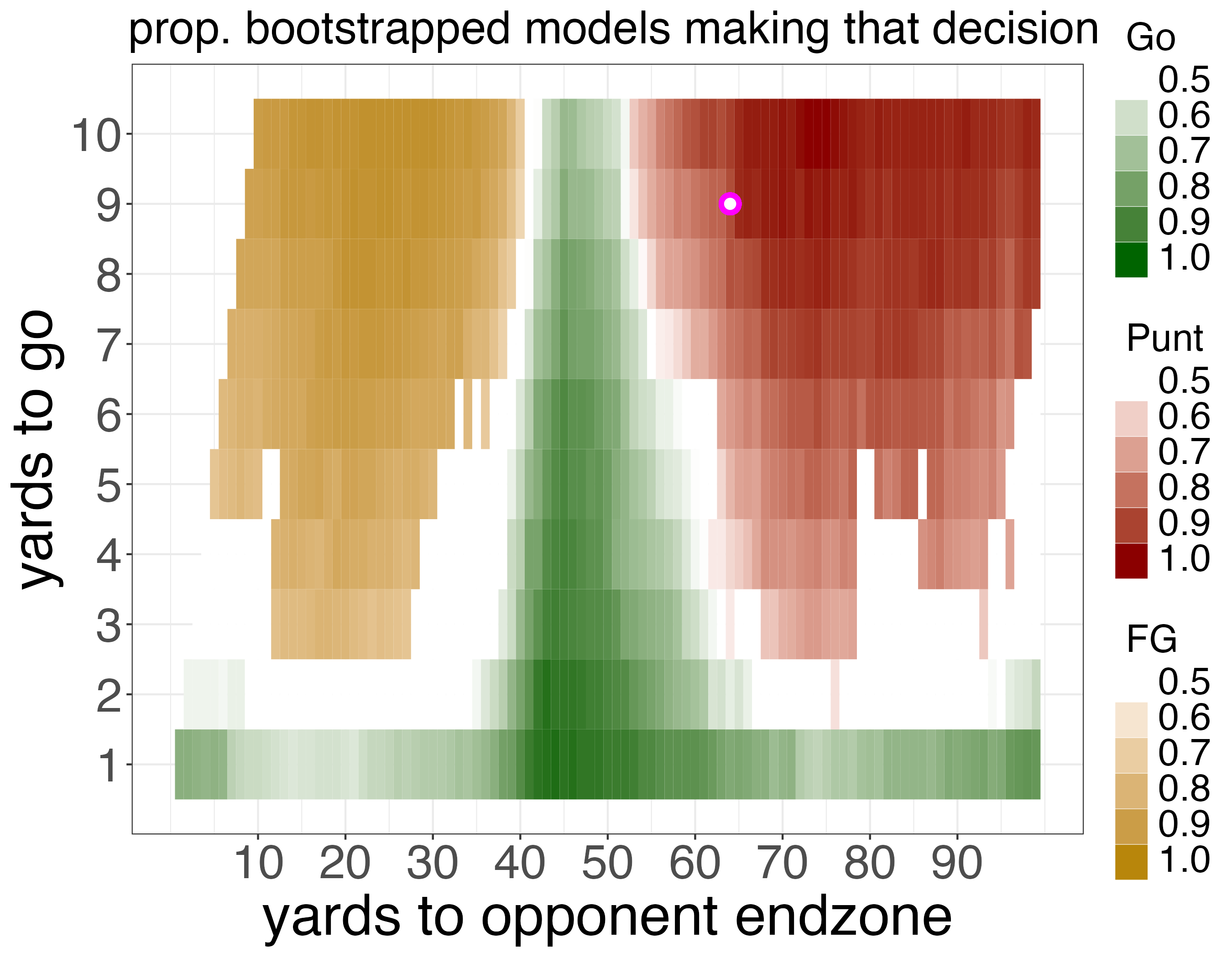}
    \end{minipage}
    \caption{
        Decision charts for example play 4.
    }
    \label{fig:ex_play_4}
\end{figure}

\textbf{Example play 4.} In Figure~\ref{fig:ex_play_4} we visualize the decision procedure for a fourth-down play in which the Ravens had the ball against the Jets in Week 1 of 2022.
$\punt$ provides a slight edge over $\go$ according to the $\wp$ point estimate ($+1.3\%$ $\wp$).
But most of the bootstrapped models say $\punt$ is better and the $90\%$ confidence interval of the estimated gain in win probability by punting is $[0\%, 3\%]$, which is positive.
So, even if the edge is small, we are confident in this edge and recommend that the Ravens should $\punt$.
Further, in the bottom right plot, notice how most of the colors are dark outside of a large white boundary region.
This indicates that we have higher certainty in the estimated optimal decision at this game-state.

\begin{figure}[hbt!]
    \centering
    \includegraphics[width=1\textwidth]{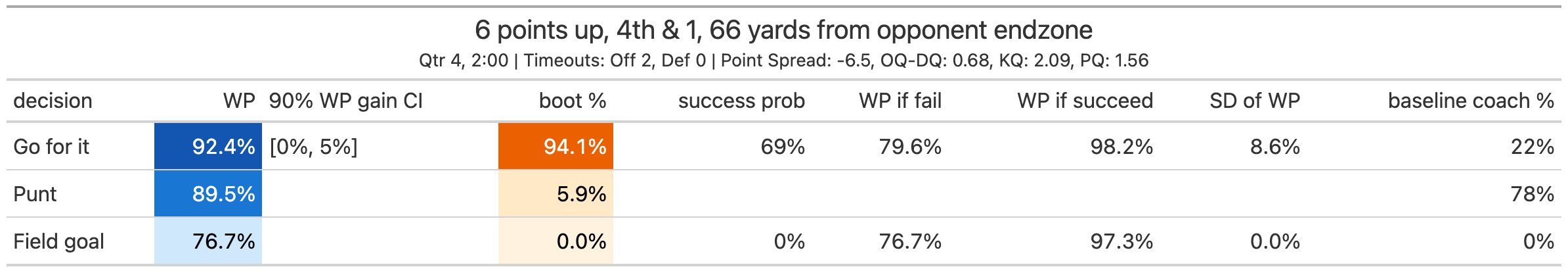}
    \begin{minipage}{\textwidth}
        \centering
        \includegraphics[width=0.48\textwidth]{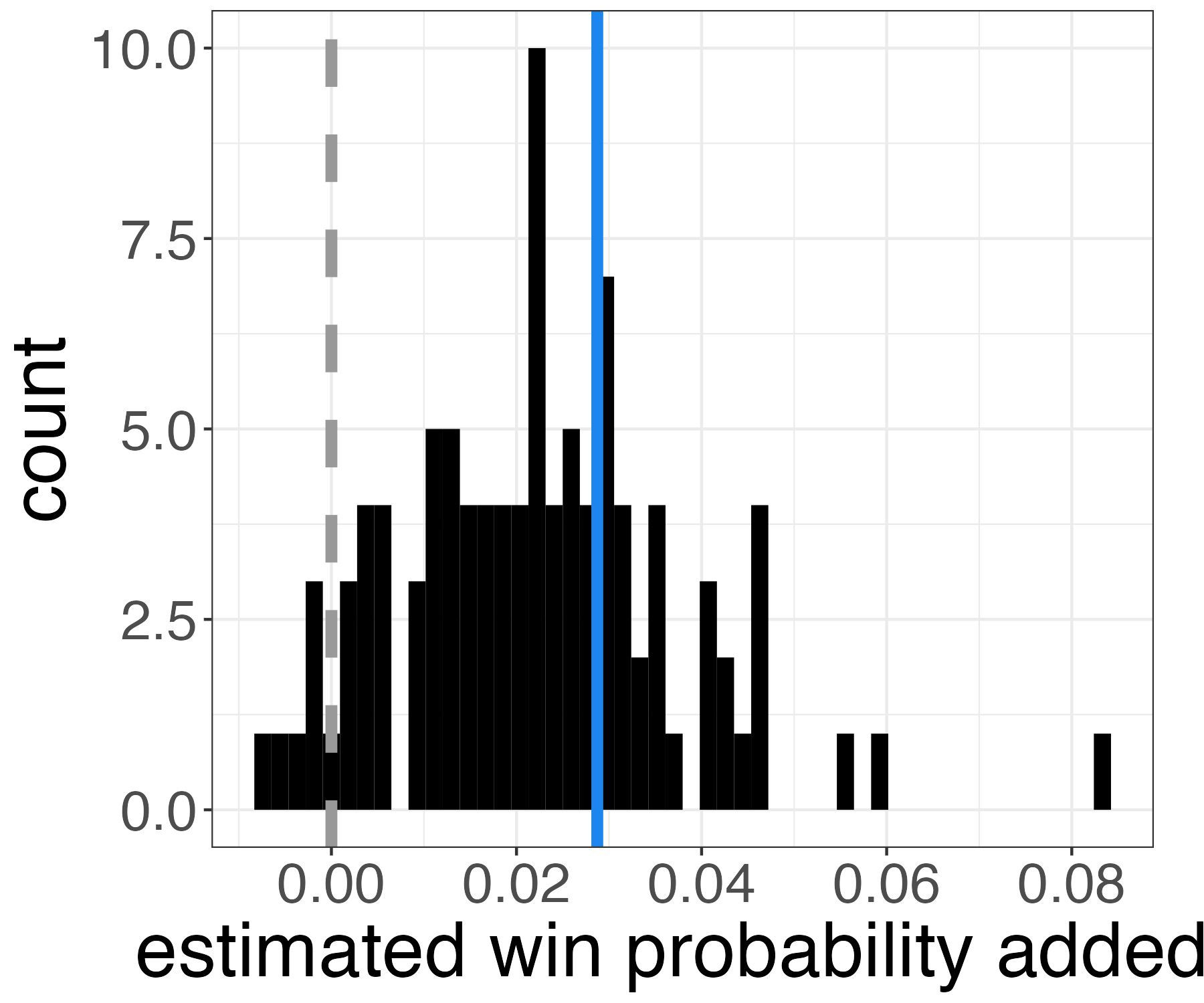}
    \end{minipage}
    \begin{minipage}{\textwidth}
        \centering
        \includegraphics[width=0.48\textwidth]{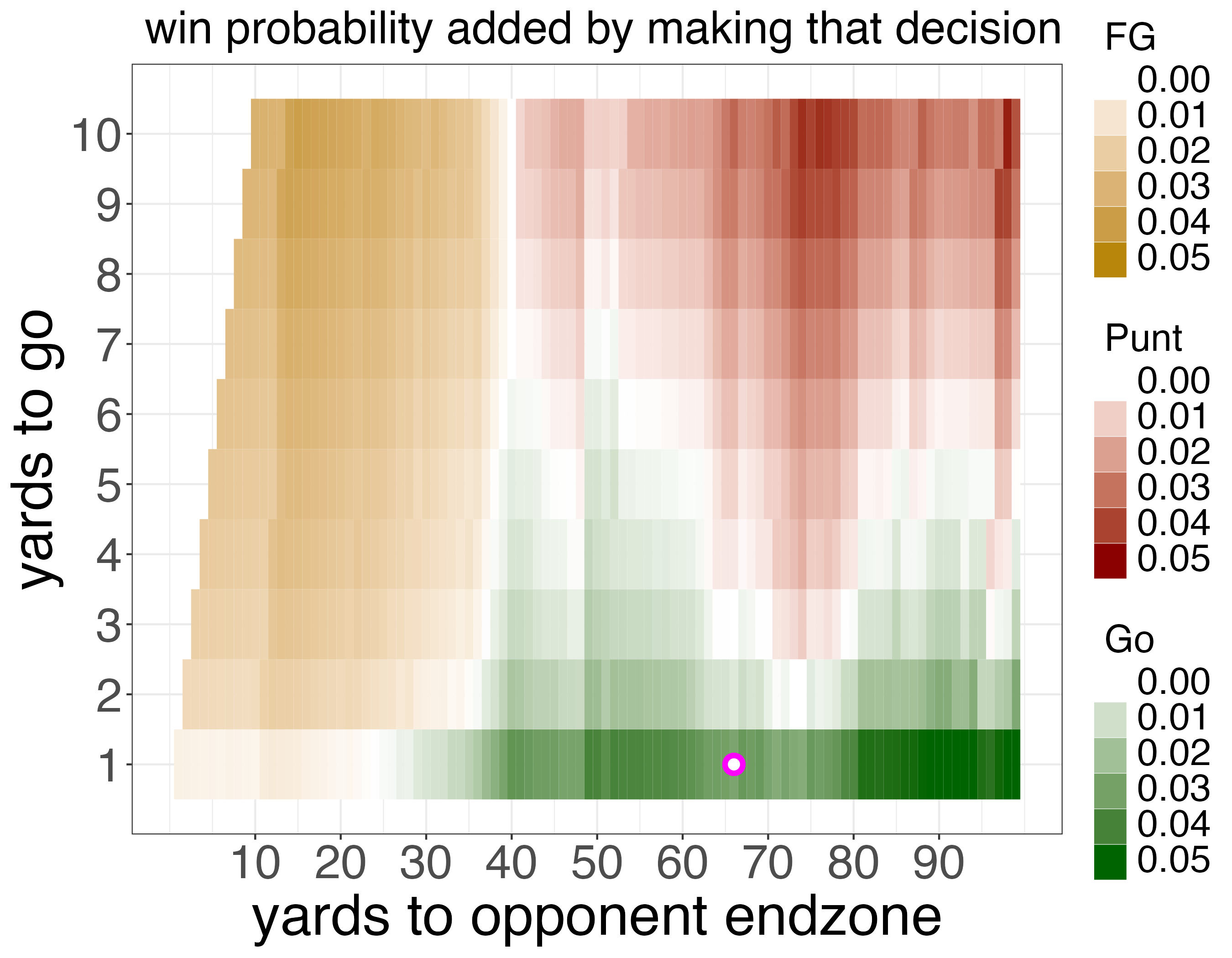}
        \includegraphics[width=0.48\textwidth]{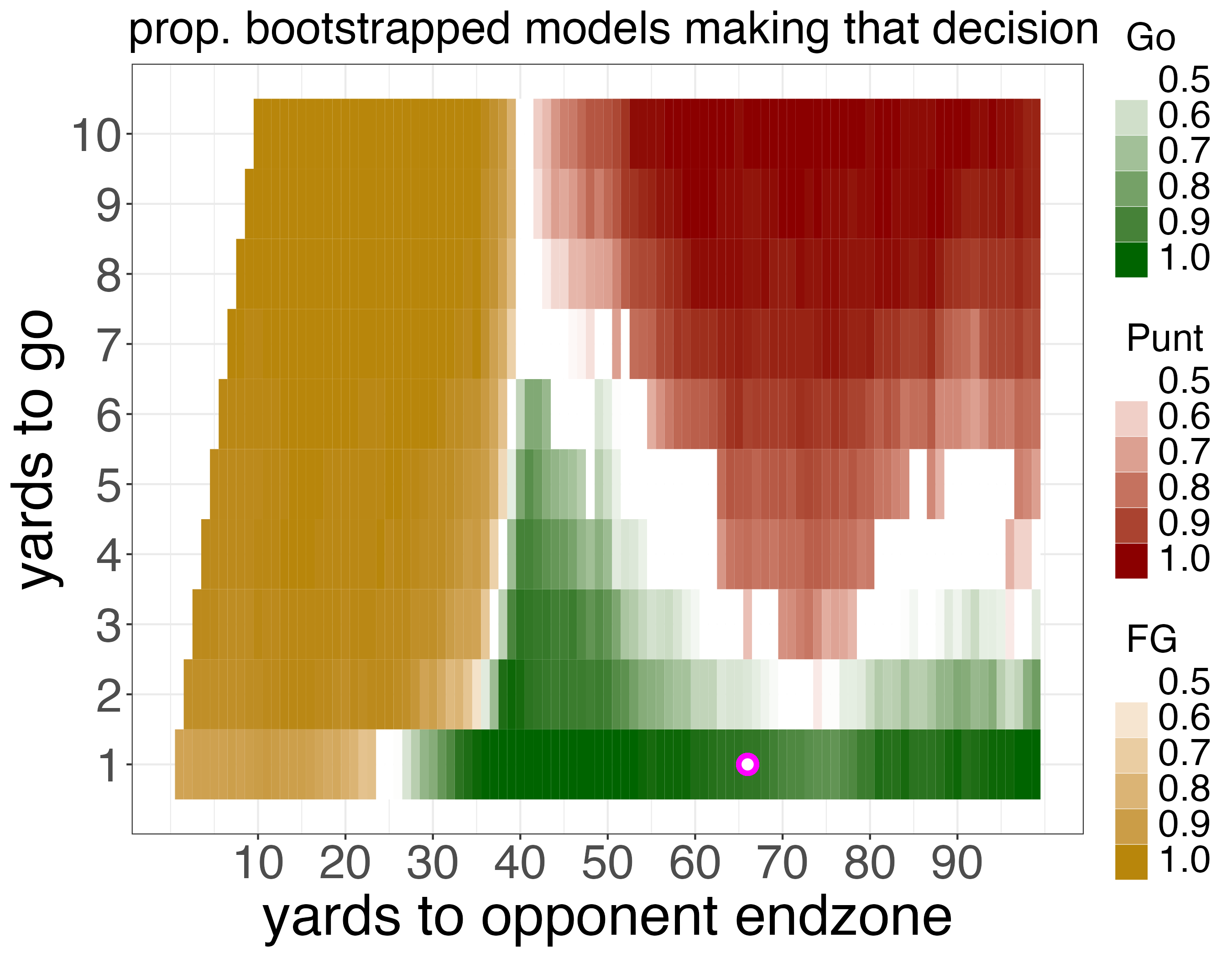}
    \end{minipage}
    \caption{
        Decision charts for example play 5.
    }
    \label{fig:ex_play_5}
\end{figure}

\textbf{Example play 5.} In Figure~\ref{fig:ex_play_5} we visualize the decision procedure for an infamous fourth-down play in which the Raiders had the ball against the Rams in Week 14 of 2022.
$\go$ provides a strong edge over $\punt$ according to the $\wp$ point estimate ($+2.9\%$ $\wp$).
Further, many of the bootstrapped models say $\go$ is better and the $90\%$ confidence interval of the estimated gain in win probability by going for it is $[0\%, 5\%]$, which is positive.
Thus, we recommend that the Raiders should $\go$.\footnote{
    In real life, the Raiders punted. Then, Rams quarterback Baker Mayfield countered with a successful 98-yard drive to win the game. 
}

\section{The difficulty of formulating a probabilistic state-space win probability model}\label{app:stateSpaceModelBias}

Proprietors of probabilistic models believe that introducing bias in order to reduce variance improves the overall accuracy of the resulting win probability estimator.
Nevertheless, these models are subject to their own set of issues, and we believe they aren't as low-variance as some analysts claim.
Properly modeling the distribution of the outcome of a play is nontrivial.
In contrast to the simple binary win/loss outcome of statistical win probability models, the outcome variable of a transition probability model is the next game-state, which could include a change in yardline, score, time, or timeouts.
This distribution is quite complex: there is a spike at gaining 0 yards for incompletions, a spike for a touchdown, spikes for penalties, and other smooth possibly multimodal distributions for the outcome of run or pass plays, each of which change as a function of team quality and other confounders \citet{BiroRunDensity,BiroPassDensity}.
Typical transition probability models are riddled with selection bias, as a coach generally chooses play calls that work for his specific players and don't generalize to an ``average'' team.
Further, uncertainty in these transition probabilities may, after being properly propagated through a state-space model, result in just as much (if not more) uncertainty in estimated win probability than estimates from statistical models.
Additionally, one must take great care to carefully encode all the subtle rules of football into her model, and one needs sufficient computing power to simulate enough games to estimate win probability with enough granularity.
We look forward to a public facing exploration of probabilistic win probability models in the future.

\end{document}